
\input phyzzx.tex


\def\np{Nucl. Phys.}
\def\pl{Phys. Lett.}

\def\prl{Phys. Rev. Lett.}
\def\cmp{Comm. Math. Phys.}
\def\intmp{Intern. J. Mod. Phys.}
\def\mpl{Mod. Phys. Lett.}
\def\dd{\hbox{d}}


\Pubnum={UU-HEP-92/6}
\date{June, 1992 }

\titlepage
\title{\bf The Complete Structure of the Cohomology Ring and Associated
Symmetries in $D=2$ String Theory}

\author{Yong-Shi Wu}
\address{Department of Physics, University of Utah
\break Salt Lake City, Utah 84112, USA}
\author{Chuan-Jie Zhu}
\address{Department of Physics, University of Utah
\break Salt Lake City, Utah 84112, USA \break
{\rm and} \break International School for Advanced Studies (SISSA/ISAS)
\break {\rm and} \break INFN, Sezione di Trieste \break
via Beirut 2-4, I-34013 Trieste, Italy\footnote*{\rm Present address.} }

\vglue .5cm

\abstract\nobreak
{We determine explicitly all structure constants of the whole
chiral BRST cohomology ring in $D=2$ string theory including both
the discrete states and tachyon states. This is made possible by
establishing several identities for Schur polynomials with operator
argument and exploring associativity. Furthermore we find that
the (chiral) symmetry algebra of the charges obtained by using the descent
equations can actually be read off from the cohomology
ring structure by simple operation involving the ghost field $b$.
We also determine the enlarged symmetry algebra which contains the
charges having ghost number $-1$ and $1$. Finally the complete symmetry
transformation rules are derived for closed string discrete states by
carefully combining the left and right sectors. It turns out that
the new states introduced recently by Witten and Zwiebach are naturally
created when symmetries act on the old states.}

\vfill\eject
\endpage

\centerline{\bf 1. Introduction}

\REF\GMA{D. J. Gross and A. A. Migdal, \prl\ {\bf 64} (1990) 717. }

\REF\GMB{M. Douglas and S. Shenker, \np\ {\bf B335} (1990) 635. }

\REF\GMC{E. Brezin and V. Kazakov, \pl\ {\bf B236} (1990) 144. }

\REF\KMA{M. Douglas, \pl\ {\bf B238} (1990) 176. }

\REF\KMB{E. Witten, \np\ {\bf B340} (1990) 281. }

\REF\PA{ A. M. Polyakov, \pl\  {\bf B180} (1981) 207.}

\REF\GTW{ A. Gupta, S. Trivedi and M. Wise, \np\ {\bf B340} (1990) 475.}

\REF\GL{ M. Goulian and M. Li, \prl\ {\bf 66} (1991) 2051.}

\REF\KITA{ Y. Kitazawa, \pl\ {\bf B265} (1991) 262; \intmp\ {\bf A7}
(1992) 3403. }

\REF\DOT{ Vl. S. Dotsenko, \mpl\ {\bf A6} (1991) 3601. }

In the last three years or so we have witnessed a spectacular success
in exactly solving certain $D\leq 2$ string theories by exploring
first matrix models [\GMA, \GMB, \GMC], and then integrable models
[\KMA] and topological
field theory [\KMB]. However, many of the results obtained in these
approaches, though explicit and exact, can hardly be given a transparent
physical interpretation. A consensus of opinion in this field has been
that by understanding these results in the more conventional Liouville
theory approach [\PA], we may be able to gain better understanding of the
underlying physics, and to develop new methods and derive new results.
Recent achievements in this regard include: 1) exact computation
of some correlation functions in Liouville theory and explicit verification
of the agreement with matrix models [\GTW, \GL, \KITA, \DOT]; 2) better
understanding of the physical states and their symmetries in the Liouville
theory for the $D\leq 2$ string theories (see below for references). In this
paper we will report the progress we have made in the second direction.

\REF\KAC{J. Goldstone (unpublished); V. G. Kac, in { \it Group Theoretical
Methods in Physics}, Lecture Notes in Physics, vol 94.}

\REF\GKN{D. J. Gross, I. R. Klebanov and M. Newman, \np\ {\bf B350}
(1991) 621. }

\REF\PAA{A. M. Polyakov, \mpl\ {\bf A6} (1991) 635.}

\REF\APA{A. M. Polyakov, ``Singular States in 2D Quantum Gravity,'' preprint
PUPT-1289 (Sept., 1991).}

\REF\LZ{B. Lian and G. Zuckerman, \pl\ {\bf B254} (1991) 417; {\bf B266}
(1991) 21. }

\REF\BMP{
P. Bouwknegt, J. McCarthy and K. Pilch, \cmp\ {\bf 145} (1992) 541. }

\REF\BMPA{
P. Bouwknegt, J. McCarthy and K. Pilch, ``BRST Analysis of Physical
States for $2D$ (Super) Gravity Coupled to (Super) Conformal Matter,''
preprint CERN-TH-6279/91 (October, 1991). }

\REF\MM{S. Mukherji, S. Mukhi and A. Sen, \pl\ {\bf B266} (1991) 337. }

\REF\IO{K. Itok and N. Ohta, ``BRST Cohomology and Physical States in 2D
Supergravity Coupled to $\hat{c}\le 1$ Matter,'' preprint
FERMILAB-PUB-91/228-T, OS-GE 20-91, Brown-HET-834 (Sept., 1991). }

\REF\WITA{E. Witten, \np\ {\bf B373} (1992) 187. }

\REF\KMS{D. Kutasov, E. Martinec and N. Seiberg, \pl\ {\bf B276} (1992) 437. }

\REF\WITZ{E. Witten and B. Zwiebach, \np\ {\bf B377} (1992) 55.  }

Among the known soluble string theories the $D=2$ model is interesting
in many aspects: It has a simple two-dimensional space-time
interpretation, incorporates the interactions between gravity and a
massless scalar ``tachyon" field, and embraces intriguing space-time
physics such as the formation and evaporation of black holes. A
by-now-well-known feature of the $D=2$ string theory is the existence
of the unusual ``discrete states", in addition to the familiar tachyons.
Some discrete states were known [\KAC] for some time. Recently they
became popular because of their appearance first in the matrix-model
calculations of Gross, Klebanov, and Newman [\GKN] and then in the
Liouville-theory analysis by Polyakov [\PAA, \APA]. The subsequent rigorous
BRST analysis [\LZ, \BMP, \BMPA] (see also [\MM, \IO]) has indicated a
more complex pattern of the discrete states than one had expected.
Besides the states of conventional ghost number one, there are also
physical states of ghost number zero and two, appearing legitimately
as non-trivial BRST cohomology classes. The study of the discrete states
and the symmetries associated to them in $D=2$ string theory could be
a good starting point for discovering the fully fledged stringy symmetries
and formulating a background independent string theory. Moreover such a study,
combined with relevant Ward identities, could lead to systematic evaluation
of correlation functions in string theory in the Liouville theory approach.

It was Witten [\WITA] who first pointed out that the ghost number zero
states generate a ``ground ring'', which is characteristic of $D=2$ string
theory. By simple considerations involving the ground ring, he was able to
explain many aspects of the free fermion description that comes from the
matrix model (see also [\KMS]). Further developments along this line of
thought have been made in [\WITZ], where a careful treatment of closed
string BRST cohomology is given and a systematic construction of symmetry
charges from the discrete states via the ``descent equations" is proposed.
One of the main points in these works is that many of the structures they
found (at the $SU(2)$ radius) can be naturally described in terms of the
differential geometry of a certain three-dimensional cone, on which the ring
of functions is isomorphic to the ground ring of the ghost-number-zero
states.

In this paper we want to study the same problems, the discrete states and
associated symmetries in $D=2$ string theory, from somewhat different point
of view, purely algebraic and more explicit. We are motivated by
the observation that all the discrete states form a (graded) ring [\WITZ]
in the following sense in terms of BRST cohomology [\WITA]. Let $V_1(z)$
and $V_2(z)$ be two (chiral) BRST invariant operators. The operator product
$V_1(z)V_2(w)$ is also BRST invariant, so all the terms in its short distance
expansion for $z\to w$ are BRST invariant. Negative powers of $z-w$ may arise
in this short distance expansion, but the operators multiplying the negative
powers of $z-w$ are of negative dimension and thus must be BRST commutators
(as there is no nontrivial BRST cohomology class at negative dimensions).
Therefore, modulo the BRST commutators, the short distance limit of
$V_1(z)V_2(w)$ is some BRST invariant operator $V_3(w)$ (which may be zero):
$$V_1(z)V_2(w) \sim V_3(w) + \{ Q, \cdots\} . \eqn\qaq$$
This gives rise to the desired multiplication law, $V_1 \cdot V_2 = V_3$.
This procedure obviously defines an associative ring structure for the
set of all BRST cohomology classes, which is graded by ghost number.
Clearly the ground ring is only a subring of this ring, which we will call
the (chiral) BRST cohomology ring.

Instead of expressing the structure of the whole cohomology ring in terms
of the ground (sub)ring in differential geometric language [\WITA, \WITZ],
we want to directly attack the problem of explicitly calculating all the
structure constants of the whole chiral BRST cohomology ring in $D=2$
string theory, including both the discrete and tachyon states. This is
possible, since an explicit representation for all discrete states is
available in the literature [\BMPA], which involves Schur polynomials
with operator argument. Starting from this representation and using
several identities we have derived to rearrange Schur polynomials, we
found that the BRST cohomology ring is computable and have obtained explicit
results for the structure constants. We have been able to include the
tachyon states in our computation and find that the product of two negative
parity tachyon states can give rise to a discrete state of ghost number two,
or $P_{j, m}$ in the notation of [\WITZ]. This again shows the
incompleteness of the physical states of standard ghost number one.

With the structure constants of the (chiral) BRST cohomology ring
explicitly known, it is not too hard to derive the structure constants
for the associated (chiral) symmetry charge algebra or the transformation
rules for the discrete states. In doing so, the general procedure, proposed
in ref. [\WITZ], of constructing conserved charges from the discrete states
via descent equations is extremely powerful. Our results show that the
(chiral) symmetry charge algebra can be read off from the cohomology ring
structure by simple operation involving the ghost field $b$. In combining
the right and left (chiral) cohomology classes to obtain closed string
BRST cohomology, we have to pay attention to the subtleties pointed out
by Witten and Zwiebach [\WITZ], which allow the existence of more closed
string discrete states and associated symmetries than what had been
recognized before from naive combination of chiral BRST cohomologies.
With these subtleties taken into account, we have determined the
transformation rules in closed string theory. The rules obtained show
that the new states introduced by Witten and Zwiebach [\WITZ] are
naturally created when symmetry charges act on the old states.
Since the cohomology ring structure
implies the charge algebra but the converse is not true, we can say that
the cohomology ring is more fundamental than the symmetry charge algebra.

This paper is organized as follows. In the next section we review
the previous results about the classification of physical
states in the $D=2$ string theory at the $SU(2)$ point (with a vanishing
cosmological constant). In particular we present an explicit
representation of all the (chiral) discrete states in terms of BRST
invariant local operators [\WITA, \WITZ]. A very useful rearrangement
formula for $P_{j, m}$ is announced here and its proof will be deferred
to sec. 3.

\REF\PAK{I. Klebanov and A. M. Polyakov, \mpl\ {\bf A6} (1991) 3273. }

\REF\LI{ M. Li, ``Correlators of Special States in $c= 1$ Liouville Theory'',
preprint UCSBTH-91-47 (October, 1991).}

\REF\TAN{Y. Matsumura, N. Sakai and Y. Tanii, ``Coupling of Tachyons and
Discrete States in $c=1$ 2-D Gravity,'' preprint TIT/HE-186 or STUPP-92-124
(Jan., 1992); ``Interaction of Tachyons and Discrete States in $c=1$ 2-D
Quantum Gravity,'' preprint TIT/HE-187 or STUPP-124-125 (Jan., 1992).  }

In sec. 3 we explicitly determine all structure constants of the chiral BRST
cohomology ring. Part of the ring structure was already known [\WITA,
\PAK, \LI]. The new results are derived by using several useful identities
involving Schur polynomials with operator arguments, which are established
by us and proved in appendix A, and by exploring associativity. In this
section we will present some sample calculations and give the proof
of the rearrangement formula for $P_{j, m}$. Our new results include
the structure constants involving higher ghost number states, like
$O_{j_1m_1}P_{j_2m_2}$ and $Y^+_{j_1m_1}P_{j_2m_2}$ in the notation of
[\WITZ]. In particular, by explicit calculation we find
$Y^-_{j_1 m_1} Y^-_{j_2 m_2}$ is zero.

The complete chiral cohomology ring involves also the tachyon states. The
structure constants involving tachyon states are derived in sec. 4
and are compared with previous results [\TAN] about the OPE and couplings
of tachyons and discrete states in $D=2$ string theory. Amazingly we find
$T^{(-)}_pT^{(-)}_q$ is generally non-zero. Of particular interest is that
this product gives rise to a term proportional to $P_{jm}$ in addition
to the usual term $aY^-_{j+1, m}$.

In sec. 5 we first explain how one can construct conserved charges from
BRST invariant local operators. The procedure is just part of the general
procedure of constructing conserved charges via descent equations
(taking only one of the chiral sectors). Then we derive some general
formulas of getting transformation rules from the ring multiplication
laws. The complete enlarged (chiral) charge
algebra is explicitly computed.

In sec. 6 we derive the symmetry transformation rules for closed
string discrete states by combining the left and right sectors and using
the results of sec. 5. Here the relevant point is to show that the new
closed string states introduced by Witten and Zwiebach [\WITZ] through
the semi-relative cohomology are naturally created when symmetries act
on the old states. The sample calculations presented in this section
suggest some very simple relations involving the multiplication of the
operator $(a+\bar{a})$. The proof of these relations and also some
details about the calculations are presented in appendix C.

For the convenience of reader, this paper contains three appendices:
Appendix A collects relevant formulas for Schur polynomials and presents
the proof for our newly established identities which are useful in the
text. Appendix B presents the complete chiral transformation rules and the
chiral symmetry algebra by using the  chiral BRST cohomology ring structure
equations (sec. 3) and the general formulas derived in sec. 5.
Appendic C gives  some  details for the calculations of the
transformation rules in closed string theory and also the proof of two
general results in this regard in sec. 6.

\vfill\eject

\centerline{\bf 2. Representation of the Chiral Discrete States}

Like in refs. [\WITA, \WITZ], we consider the $D=2$ string theory at the
$SU(2)$ radius with vanishing world-sheet cosmological constant. We
will follow the notation of [\WITA, \WITZ] as much as possible. The
basic fields appearing in the construction of the physical states are
the world-sheet ``matter" field $X$, the Liouville field $\phi$
and the ghost fields $b$ and $c$. At the $SU(2)$ radius and vanishing
world-sheet cosmological
constant, both the fields $X$ and $\phi$ are free fields, and the
right and left movers are decoupled. Thus we may start with one (say,
the right one) of the chiral sectors. The OPE's for the basic fields are
$$\eqalign{
& X(z)X(w) \sim -\ln (z-w), \cr
& \phi(z) \phi(w) \sim -\ln (z-w), \cr
& b(z)c(w) \sim { 1\over z-w} .
} \eqn\xpb$$
The stress energy tensors are
$$ \eqalign{
& T_X = -{ 1\over 2} (\partial_z X)^2, \cr
& T_{\phi} = -{ 1\over 2} (\partial_z \phi)^2 +\sqrt{2} \partial^2_z\phi, \cr
& T_{bc} = 2 (\partial_z c) b + c\partial_z b.
} \eqn\str$$
In the following we also use the light-cone fields $X^{\pm} ={1\over \sqrt{2}}
(X\pm i\phi)$. The combined matter-Liouville stress energy tensor is
$$ T = T_X + T_{\phi} = -\partial_z X^+ \partial_z X^- - i(\partial^2_z X^+
 - \partial^2_z X^-) . \eqn\tt$$
The central charge of $T$ is 26 and we have the following nilpotent  BRST
charge
$$Q = \oint_0 [\dd z] :c(z) (T(z) + \partial_z c(z) b(z)) :  , \eqn\qq$$
where $[\dd z] \equiv { \dd z \over 2\pi i }$. Here and in what follows the
normal ordering for the ghost fields is such that the following is satisfied:
$$ b(z) c(w) = { 1\over z-w} + :b(z) c(w):  . \eqn\norbc$$

According to the general principle of BRST quantization, the physical states
are defined as $Q$-closed but not $Q$-exact states, i.e. $Q$-cohomology
classes.
By a one to one correspondence between states and fields, one can also define
the states in terms of local field operators. We will use the local field
operator representation exclusively in this paper so that a physical state is
represented as a BRST invariant but not exact local operator.

Before presenting the explicit representation for all the discrete states, a
technical remark is in order. It is easy to see that from the matter field
$X(z)$ one can construct an $SU(2)$ Kac-Moody algebra:
$$\eqalign{
J_{\pm}(z) = e^{\pm i\sqrt{2} X(z) } ,~~~
J_3(z) = { i\over \sqrt{ 2} } \partial_z X(z),
} \eqn\kac$$
with the following OPE's
$$\eqalign{
& J_+(z) J_-(w) \sim { 1\over (z-w)^2} + { 2\over z-w} J_3(w) ,\cr
& J_3(z) J_{\pm}(w) \sim \pm { 1\over z-w} J_{\pm} (w) , \cr
& J_3(z) J_3(w) \sim { 1/2 \over (z-w)^2} ,
} \eqn\kacc$$
Of particular importance is the fact that the zero modes of $J_i$,
$$\hat{J} _{\pm, 3} =\oint_0  [\dd z] J_{\pm, 3}(z), \eqn\zero$$
commutes with $Q$: $[Q, \hat{J}_{\pm, 3} ] = 0$. Thus all
the physical states should form $SU(2)$ multiplets. In fact all the discrete
physical states do fall into  finite dimensional multiplets. If there exists
a physical states $V_{j, j}(w)$ such that
$$\eqalign{
& [\hat{J}_+, V_{j, j} (w)] = 0, \cr
& [\hat{J}_3, V_{j, j} (w)] = j V_{j, j} (w),
} \eqn\deffj$$
one can get the remaining  states in the multiplet by using the
$\hat{J}_-$ operator
$$ \eqalign{
V_{j, m} (w) & =
\left( { (j+m)!\over (2j)!(j-m)!} \right)^{1/2}
\underbrace{[\hat{J}_-, [\hat{J}_-, \cdots, [\hat{J}_-}_{j-m},
V_{j, j}(w)\underbrace{]\cdots ]}_{j-m}\cr
&\equiv
\left( { (j+m)!\over (2j)!(j-m)!} \right)^{1/2}
(\hat{\hat{J}}_-)^{j-m} V_{j, j}(w), } \eqn\ddii$$
where
$$ \hat{\hat{J}}_- V_{j, m-1} =\oint_w[\dd z] J_-(z) V_{j, m} (w).
\eqn\raise$$

After all these preliminaries we can now summarize the results of
[\LZ, \BMP, \BMPA] in compact form. All the physical states are
divided into two categories: relative physical states that are
annihilated by $b_0$ and absolute physical states that are not
annihilated by $b_0$. Apart from the tachyon states (see sec. 4),
the other relative physical states have discrete momenta $(p_X,
p_\phi)$ and are created by following local vertex operators:
$$\eqalign{
& O_{j, m} =
\left( { (j+m)!\over (2j)!(j-m)!} \right)^{1/2}
(\hat{\hat{J}}_-)^{j-m} O_{j, j} , \cr
& O_{j, j} = :\left( -S_{2j}(-iX^-) + \sum_{q=1}^{2j} S_{2j-q}(-iX^-)
{ c\partial^{q-1} b \over (q-1)! } \right) e^{ 2ijX^+} :, \cr
& Y^{\pm}_{j, m}  = c:
\left( { (j+m)!\over (2j)!(j-m)!} \right)^{1/2}
(\hat{\hat{J}}_-)^{j-m}
e^{ (ij X +(1\mp j)\phi)\sqrt{2} } :, \cr
& P_{j, m} =
\left( { (j+m)!\over (2j)!(j-m)!} \right)^{1/2}
(\hat{\hat{J}}_-)^{j-m} P_{j, j} , \cr
& P_{j, j} =
\sum_{q = 1}^{2j + 1}: S_{2j+1-q}(-iX^+ - iX^-) {\partial^{q+1}
c\over (q+1)! } ce^{ -2iX^+ +2i(j+1)X^-}:  ,
} \eqn\res$$
where $m= -j, -j+1, \cdots, j$ and $j = 0, 1/2, 1, 3/2, \cdots$.
In the above equation $S_k(-iX^-)$ is a Schur polynomial of degree $k$
with the operator $-iX^-$ as its argument. We refer the reader to
appendix A for our notation and relevant properties of the Schur
polynomials. In Fig. 1 we show the $(p_X, p_\phi)$ momenta of these
discrete states. According to the type of Liouville dressing,
$O_{j,m}$ and $Y^+_{j,m}$ are called the ``plus'' states,  and
$Y^-_{j,m}$ and $P_{j,m}$ are called the ``minus'' states. Note that
the states $O_{j,m}$, $Y^{\pm}_{j,m}$ and $P_{j,m}$ have ghost number
$0$, $1$ and  $2$ respectively.

A remark is in order about the representation of $P_{j, j}$ (and all the
$P_{j, m}$'s by $SU(2)$ action). According to [\BMP], the non-triviality of
$P_{j, j}$ could be proved by changing the prefactor to a representation
in terms of only the light-cone variable $X^+$ by adding some BRST exact
terms. We find that it is quite difficult to prove the
non-triviality of $P_{j, j}$ with the representation given in $\res$.
In the process of seeking a general procedure of handling this rearrangement
problem we have found the following general representation for $P_{j, j}$:
$$ \eqalign{
P_{j, j} & \equiv
\sum_{q = 1}^{2j + 1} S_{2j+1-q}(-iX^+ - iX^-) { \partial^{q+1}
c\over (q+1)! }c e^{ -2iX^+ +2i(j+1)X^-}   \cr
& = { \Gamma(2j+2)\Gamma(\delta+1) \over \Gamma(2j+1+\delta) }
\sum_{q = 1}^{2j + 1} S_{2j+1-q}(-iX^+ - i\delta X^-) { \partial^{q+1}
c\over (q+1)! } ce^{ -2iX^+ +2i(j+1)X^-} \cr
& \equiv { \Gamma(2j+2)\Gamma(\delta+1) \over \Gamma(2j+1+\delta) }
 P_{j, j}(\delta) .  } \eqn\newf$$
with $\delta$ an arbitrary real number. Later we will see the usefulness of
this formula. It will be proved in the next section.

As for the absolute physical states, they can be obtained from the relative
physical states by ``multiplying'' the following local operator [\WITZ, \LI]
$$ a = {1\over \sqrt{2} } [Q, \phi] = \partial c +{1\over \sqrt{2} }c
\partial \phi, \eqn\aaa$$
In the usual sense $\phi$ is not a conformal field, but $a$ is. Obviously
$a$ is BRST invariant and is not BRST exact in the usual space of conformal
fields. This being so,
we can get absolute physical states from the relative ones by
forming the local product  with $a$, i.e.
$$(aV_{jm})(w) = \oint_w[\dd z] { 1\over z-w} a(z) V_{jm}(w). \eqn\defab$$
The local multiplication with $a$ commutes with the $SU(2)$ action.
Thus, $(aV_{j,m})$, or simply $aV_{j,m}$ when there is no confusion,
has the same $SU(2)$ quantum numbers, as well as the same $(p_X, p_\phi)$
momenta, as $V_{j, m}$. This way of forming an absolute physical state
can be viewed as the multiplication law for the BRST invariant operator $a$
and a relative physical state in the ring. This makes it easy to derive
the ring multiplication involving absolute physical states from that of
relative physical operators, as exemplified in the following:
$$\eqalign{
& \qquad O_{j_1m_1} O_{j_2m_2} = O_{j_1+ j_2, m_1+ m_2} \to \cr
& aO_{j_1m_1} O_{j_2m_2} = O_{j_1m_1} aO_{j_2m_2}
= aO_{j_1+ j_2, m_1+ m_2}.}
\eqn\oooa$$
Note that the multiplication of $a$ with itself gives zero. So only the
ring structure of the relative physical states will be computed.

\vfill
\eject

\centerline{ \bf 3. Cohomology Ring of the Discrete Chiral States}

First let us summarize the known results on the explicit ring structure
of the discrete chiral states:

1) Witten [\WITA] derived the chiral ground ring as the ring of
polynomial functions in $x\equiv O_{1/2, 1/2}$ and $y\equiv O_{1/2, -1/2}$.
This implies that the operators $O_{jm}$ are closed under multiplication:
$$ O_{j_1m_1} O_{j_2m_2} =
\langle j_1m_1j_2m_2\mid j_1+j_2, m_1+m_2\rangle
O_{j_1+ j_2, m_1+ m_2}, \eqn\ground$$
up to normalization.  We will prove this result by explicit calculation
and find the rescaling of $O_{jm}$ such that all the structure
constants here become unity.( Another, perhaps more natural, normalization
will be presented at the end of this section.)

2) Polyakov and Klebanov [\PAA] derived the following OPE
($Y^{\pm}_{jm} = c\Psi^{(\pm)}_{jm}$):
$$\eqalign{
& \Psi^{(+)}_{j_1m_1}(z) \Psi^{(+)}_{j_2m_2}(w) \sim \cdots + { 1\over z-w}
(j_2m_1-j_1m_2)\Psi^{(+)}_{j_1+j_2-1, m_1+m_2}(w), \cr
&  \Psi^{(-)}_{j_1m_1}(z) \Psi^{(-)}_{j_2m_2}(w) \sim \cdots + { 1\over z-w}
\times 0, \cr
&   \Psi^{(+)}_{j_1m_1}(z) \Psi^{(-)}_{j_2m_2}(w) \sim \cdots + { 1\over z-w}
\times 0, \qquad j_1\ge j_2+1, \cr
& \Psi^{(+)}_{j_1m_1}(z) \Psi^{(-)}_{j_1+j_2-1, -m_1-m_2}(w) \sim \cdots -
{ 1\over z-w} (j_2m_1-j_1m_2)\Psi^{(-)}_{j_2, -m_2}(w).
} \eqn\poly$$
This gives a term $aY^{\pm}_{j_2\pm(j_1-1), m_1+m_2}$ in the product
$Y^{\pm}_{j_1m_1}Y^{\pm}_{j_2m_2}$. However, this is only part of the
multiplication law.  As we will see later, in addition, there is another
term proportional to $P_{j_2-j_1, m_1+m_2}$ in the product of $Y^+_{j_1m_1}
Y^-_{j_2m_2}$.  This is
due to the higher pole terms in $\poly$ which were not computed explicitly.

3) Li [\LI] obtained the following multiplication law:
$$O_{j_1m_1} Y^+_{j_2m_2} = {j_2\over j_1+j_2} Y^+_{j_1+j_2, m_1+m_2}
- {1\over j_1+j_2} (j_2m_1-j_1m_2)
aO_{j_1+j_2-1, m_1+m_2}. \eqn\lieq$$
It was obtained partly by explicit calculation and partly by using the fact
that $Y^+_{jm} $ acts on the ground ring as a differential operator [\WITA].
According to our experience the second term in $\lieq$ can be derived
from the associativity of the ring. We will explore associativity
in several places either to check our results or to derive the ring structure
constants which are not explicitly calculable.

To begin with, let us first compute $O_{j_1m_1}O_{j_2m_2}$. Counting the
momenta and ghost number, we know that this product must
be proportional to $O_{j_1+j_2, m_1+m_2}$:
$$ O_{j_1m_1}O_{j_2m_2}= g(j_1, j_2)
\langle j_1m_1j_2m_2\mid j_1+j_2, m_1+m_2\rangle
O_{j_1+j_2, m_1+m_2}, \eqn\ooo$$
where $\langle j_1m_1j_2m_2\mid j_1+j_2, m_1+m_2\rangle$ are the
Clebsch-Gordan coefficients.
To determine the unknown function $g(j_1, j_2)$, one may consider
some special values of $m_1$ and $m_2$ such that a
direct calculation is possible. Let us take $m_1=j_1$, $m_2=j_2$ and
$j_1= {1\over 2}$. The general case can be reached by induction.
We have\footnote{\S}{Some useful formulas about the OPEs
involving Schur polynomials are given in the appendix A. }
$$ O_{1/2,1/2}O_{j, j}(w) = \oint_w[\dd z] { 1\over z-w}
O_{1/2,1/2}(z) O_{j, j} (w),
\eqn\aaaa$$
with
$$\eqalign{
  O_{1/2,1/2}&(z) O_{j, j} (w)= :(c(z)b(z) +i\partial_zX^-(z)) e^{iX^+(z)} :\cr
&  \times :(-S_{2j} (-iX^-(w)) +
\sum_{q=1}^{2j}S_{2j-q}(-iX^-(w)) { c(w)\partial_w^{q-1}
b(w)\over (q-1)! } )e^{2ijX^+(w)}:   \cr
 = & :\left\{ -\left[ c(z)b(z)+i\partial_z X^-(z) + 2j{ 1\over z-w} \right]
\sum_{k=0}^{2j} S_{2j-k} (-iX^-(w)){ 1\over (z-w)^k } \right. \cr
&  + \sum_{q=1}^{2j}\left[(c(z)b(z) +i\partial_z X^-(z)+ 2j{ 1\over z-w} )
{ c(w)\partial_w^{q-1} b(w) \over (q-1)! }
\right.\cr
& \left.
+ { c(z)\partial_w^{q-1}b(w) \over (q-1)! } { 1\over z-w}
+ b(z) c(w) { 1\over (z-w)^q}
+ { 1\over (z-w)^{q+1} } \right]\cr
&  \left. \times \sum_{k=0}^{2j-q} S_{2j-q-k}
(-iX^-(w)) { 1\over (z-w)^k } \right\} e^{ iX^+(z) + 2ijX^+(w) } : .
} \eqn\expan$$
Let us compute the various terms appearing in $\expan$. First one can show
that the term containing four ghost fields is zero:\footnote*{The
exponential factor $  e^{ iX^+(z)+2ijX^+(w)}$ is suppressed.}
$$ \eqalign{
\sum_{q=1}^{2j} & :c(z)b(z) c(w) { \partial_w^{q-1} b(w) \over (q-1)!}
\sum_{k=0}^{2j-q}S_{2j-q-k} (-iX^-(w)) { 1\over (z-w)^k }
: \cr
& = \sum_{k=1}^{2j} :S_{2j-k}(-iX^-(w)) { 1\over (z-w) ^{k-1} } c(z)b(z)c(w)
\sum_{q=1}^k (z-w)^{q-1} { \partial_w^{q-1} b(w) \over (q-1)! }
: \cr
& = \sum_{k=1}^{2j} :S_{2j-k}(-iX^-(w)) { 1\over (z-w) ^{k-1} } c(z)b(z)c(w)
(b(z) + o( (z-w)^k ) ) : \cr
& = o(z-w) .
} \eqn\oko$$
The integration over $z$ then gives vanishing contribution. For the terms
containing no ghost field, we have
$$ \eqalign{
& \left[ -(i\partial_zX^-(z) + 2j{ 1\over z-w} )
\sum_{k=0}^{2j} S_{2j-k} (-iX^-(w)){ 1\over (z-w)^k }\right. \cr
&\qquad\qquad\qquad\quad
\left.+ \sum_{q=1}^{2j} \sum_{k=0}^{2j-q} S_{2j-q-k} (-iX^-(w))
{ 1\over (z-w)^{q+k-1}  } \right]
\cr
& =  \left[  \sum_{k=0}^{2j} (k-2j) S_{2j-k} (-iX^-(w)){ 1\over (z-w)^{k+1}
 } + \sum_{k=0}^{2j} { 1\over (z-w)^{k}  }P(k)  + o(z-w) \right]
 ,
} \eqn\okoa$$
where
$$P(k) = \sum_{l=k}^{2j} S_{2j-l} (-iX^-(w))
\Big(-i { \partial_w^{l-k+1}X^-(w) \over (l-k)! } \Big) . \eqn\okob$$
By using the last equation in appendix A we find that all the singular terms
in $\okoa$ are zero and the regular term $P(0)$ is $(2j+1)S_{2j+1} (-i
X^-(w) )$. A similar but a bit more tedious calculation of the remaining
terms (i.e. those containing two ghost fields) gives
$$ -(2j+1)\sum_{q=1}^{2j+1}: S_{2j+1-q} (-iX^-(w)) { c(w)\partial_w^{q-1}b(w)
\over (q-1)! }  e^{ iX^+(z) + 2ijX^+(w) } : +o(z-w). \eqn\okoc$$
It is remarkable that all singular terms in the OPE of
$O_{1/2, 1/2}(z)$ and $O_{j, j}(w)$ are zero exactly
without any BRST exact terms.

Combining the above results, we have
$$ O_{1/2, 1/2} O_{j, j} = -(2j+1) O_{j+1/2, j+1/2} . \eqn\okod$$
Using it recursively one obtains
$$ O_{j_1j_1} O_{j_2, j_2} = -{ (2(j_1+j_2))!\over (2j_1)!(2j_2)! }
O_{j_1+j_2, j_1+j_2} . \eqn\okoe$$
Recalling the explicit expression for the relevant Clebsch-Gordan coefficients
here:
$$\eqalign{
\langle j_1m_1j_2m_2&\mid jm\rangle \mid_{j=j_1+j_2, m=m_1+m_2}\cr
& = \left[
{ (2j_1)!\over (j_1+m_1)!(j_1-m_1)! }
{ (2j_2)!\over (j_2+m_2)!(j_2-m_2)! }
{ (j+m)!(j-m)! \over (2j)! } \right]^{1/2} , } \eqn\okof$$
and rescaling $O_{jm}$ to\footnote*{It is not quite natural to scale the
vertex operator $O_{jm}$'s (also $Y^{\pm}_{jm}$ and $P_{jm}$'s) by an $m$
dependent factor which spoil the $SU(2)$ structure among them. This
normalization (for $Y^{\pm}_{jm}$'s) was used in [\PAK] and is quite useful
as to make all the formulas explicit and simple. We will give the structure
equations in their $SU(2)$ covariant form at the end of this section. }
$$ -(2j)! \left[ { (j+m)!(j-m)!\over (2j)! }\right]^{1/2} O_{j, m}~~ ,
\eqn\okog$$
we get the following ring multiplication law:
$$O_{j_1m_1} O_{j_2m_2} = O_{j_1+j_2, m_1+m_2} . \eqn\okoh$$

As the second example, we derive the multiplication law for
$Y^+_{j_1m_1}Y^+_{j_2m_2}$ from the first equation of $\poly$ as follows.
First by momentum and ghost number counting we conclude that
$$Y^+_{j_1m_1}Y^+_{j_2m_2} = g(j_1, j_2)
\langle j_1m_1j_2m_2\mid j_1+j_2-1, m_1+m_2\rangle
aY^+_{j_1+j_2-1, m_1+m_2} . \eqn\poa$$
In order to determine the function $g(j_1, j_2)$ we consider the
special value $m_1=j_1$ and $m_2= j_2-1$. In this special case the higher
order pole terms in the first equation of $\poly$ vanish: Because
$Y^+_{jm}(z)=c(z)\Psi^{(+)}_{jm}(z)$, and $:c(z)c(w):\mid_{z=w}=0$,
the single pole term is the only term contributing to
$Y^+_{j_1m_1}Y^+_{j_2m_2}$. Noticing that
$aY^+_{jm} = j :\partial_zc(z)Y^+_{jm}(z):$
and rescaling $Y^+_{jm}$ as $\Psi^{(+)}_{jm}$ in [\PAA] to
$$ \sqrt{j/2} (2j-1)!\left[ { (j+m)!(j-m)!\over (2j-1)! } \right]^{1/2}
Y^+_{jm} , \eqn\poc$$
we get the following ring multiplication law by making use of $\poly$:
$$Y^+_{j_1m_1}Y^+_{j_2m_2} = {1\over j_1+j_2-1} (m_1j_2-m_2j_1)
 aY^+_{j_1+j_2-1, m_1+m_2}.\eqn\poaa$$

The other ring multiplication law involving the plus states only is
$O_{j_1m_1}Y^+_{j_2m_2}$.
By momentum and ghost number counting we have
$$\eqalign{ O_{j_1m_1}Y^+_{j_2m_2}= & g_1(j_1, j_2)
\langle j_1m_1j_2m_2\mid j_1+j_2, m_1+m_2\rangle Y^+_{j_1+j_2, m_1+m_2} \cr
& + g_2(j_1, j_2)\langle j_1m_1j_2m_2\mid j_1+j_2-1, m_1+m_2\rangle
aO_{j_1+j_2-1, m_1+m_2} .
} \eqn\poaf$$
The unknown function $g_1(j_1, j_2)$ can easily  be computed by setting
$m_1= j_1$ and $m_2=j_2$ because the second term then vanishes. After
appropriate rescaling as in $\okog$ and $\poc$ we fnd
$$O_{j_1m_1}Y^+_{j_2m_2}= {j_2\over j_1+j_2} Y^+_{j_1+j_2, m_1+m_2}
+ \tilde{g}_2(j_1,j_2)(m_1j_2-m_2j_1)aO_{j_1+j_2-1, m_1+m_2} .\eqn\xxx$$
To compute $\tilde{g}_2(j_1, j_2)$, we explore the associativity of the
ring multiplication as follows.
Multipling both sides of $\xxx$ by $Y^+_{j_2m_3}$ from the right,
$$\eqalign{
(O_{j_1m_1}Y^+_{j_2m_2})Y^+_{j_3m_3} = &{j_2\over j_1+j_2}
Y^+_{j_1+j_2, m_1+m_2}Y^+_{j_3m_3}  \cr
& + \tilde{g}_2(j_1,j_2)(m_1j_2-m_2j_1)
aO_{j_1+j_2-1, m_1+m_2}Y^+_{j_3m_3}  .
} \eqn\axxx$$
On one hand, we have
$$\eqalign{
aO_{j_1+j_2-1, m_1+m_2}Y^+_{j_3m_3}  = &
a( O_{j_1+j_2-1, m_1+m_2}Y^+_{j_3m_3}) \cr
=& { j_3 \over j_1+j_2+ j_3 -1} aY^+_{j_1+j_2+j_3-1, m_1+m_2+m_3} ,
} \eqn\axxxa$$
because of $a^2= 0$. On the other hand, for the LHS of $\axxx$ we have
$$\eqalign{
(O_{j_1m_1}Y^+_{j_2m_2})Y^+_{j_3m_3} =& O_{j_1m_1} (Y^+_{j_2m_2}
Y^+_{j_3m_3})  \cr
=& O_{j_1m_1} \Big( (m_2j_3 - m_3j_2) { 1\over j_2+ j_3 -1}
aY^+_{j_2+j_3-1, m_2+m_3} \Big) \cr
=&  (m_2j_3 - m_3j_2) { 1\over j_2+ j_3 -1} a(O_{j_1m_1} Y^+_{j_2+j_3-1,
m_2+m_3}) \cr
=&  (m_2j_3 - m_3j_2) { 1\over j_1 + j_2+ j_3 -1}
aY^+_{j_1+j_2+j_3-1, m_1+m_2+m_3} .
} \eqn\pad$$
Substituting $\axxxa$ and $\pad$ into $\axxx$ we get
$\tilde{g}_2(j_1, j_2) = - 1/ (j_1+j_2)$.

To compute the ring structure constants involving the minus states like
$Y^-_{jm}$, one can not simply use $\poly$ because of the higher order
pole terms. Their existence can be seen from the product
$Y^-_{j_1m_1}Y^-_{j_2m_2}$. By momentum and ghost number counting we have
$$ Y^-_{j_1m_1}Y^-_{j_2m_2} =g(j_1, j_2)
\langle j_1m_1j_2m_2\mid j_1+j_2, m_1+m_2\rangle P_{j_1+j_2, m_1+m_2} .
\eqn\new$$
Setting $m_1=j_1$ and $m_2=j_2$ we get
$$ \eqalign{
 Y^-_{j_1j_1} Y^-_{j_2j_2}(w)= &\oint_w[\dd z]
:c(z)e^{-iX^+(z)+i(2j_1+1)X^-(z)} :
:c(w)e^{-iX^-(w)+i(2j_2+1)X^-(w)} : \cr
 =&\sum_{q=0}^{2(j_1+j_2)+1} :{ \partial^{q+1}c \over (q+1)!} c
S_{2(j_1+j_2)+1-q}(-iX^++i(2j_1+1)X^-)  \cr
& \times e^{-2iX^++2i(j_1+j_2+1)X^- }  : .
} \eqn\newa$$
 From $\new$ we see that there should be  no $aY^-_{j_1+j_1+1, j_1+j_2}$
term in $\newa$. So the $q=0$ term in $\newa$ must be BRST equivalent to
$P_{j_1+j_2, j_1+j_2}$ (up to a proportional constant). In fact one can
prove that
$cS_{2j+1}(-iX^+-i\delta X^-) e^{-2iX^+ + 2i(j+1)X^- } $
is BRST invariant for any $\delta$. For $\delta=-1$, $-2$, $\cdots$, $-(2j+1)$,
it is actually BRST exact. We have (modulo BRST exact terms)
$$cS_{2j+1}(-iX^+-i\delta X^-) e^{-2iX^+ + 2i(j+1)X^- }  = { \Gamma(2j+2+
\delta ) \over (2j+2)! \Gamma(\delta + 1) } \sqrt{ 2(j+1)}
Y^+_{j+1, j}. \eqn\newb$$
This can easily be proved by using the raising operator $\hat{\hat{T}}_+$. The
action of $a$ on
$cS_{2j+1}(-iX^+-i\delta X^-) e^{-2iX^+ + 2i(j+1)X^- } $
gives the following:
$$\eqalign{
a\Big(  cS_{2j+1}&(-iX^+-i\delta X^-) e^{-2iX^+ + 2i(j+1)X^- }\Big) \cr
 = & \Big\{
{1\over 2}(\delta -1)
\sum_{q=1}^{2j+1} {\partial^{q+1}c\over (q+1)! } c
S_{2j+1-q}(-iX^+-i\delta X^-) \cr
&  -(j+1) \partial c c S_{2j+1}(-iX^+-i\delta X^-)
\Big\} e^{-2iX^++ 2i(j+1)X^- } .
} \eqn\newc$$
By using the above formula in $\newa$ ($\delta= -(2j_1+1)$ and $j=j_1+j_2$)
we get then
$$ \eqalign{Y^-_{j_1j_1} Y^-_{j_2j_2}  = &- {j_2\over j_1+j_2+1}
\sum_{q=1}^{2(j_1+j_2)+1} {\partial^{q+1}c\over (q+1)! } c \cr
& \times
S_{2(j_1+j_2)+1-q}(-iX^+-i(2j_1+1) X^-) e^{-2iX^++ 2i(j+1)X^- } .} \eqn\newd$$
This vanishes by making use of $\newf$. Recalling $\new$, we have
$$ Y^-_{j_1m_1}Y^-_{j_2m_2}= 0, \eqn\newe$$

There is little difficulty to extend the above calculation to determine the
ring multiplication law for $Y^+_{j_1m_1} Y^-_{j_2m_2}$. The result is
$$\eqalign{
Y^+_{j_1m_1}Y^-_{j_2m_2} = & {j_1\over j_2-j_1+1} P_{j_2-j_1, m_1+m_2} \cr
& +  { 1\over j_2-j_1+1} (m_1j_2+m_2j_1+m_1) aY^-_{j_2-j_1+1, m_1+m_2},
} \eqn\xaa$$
if we rescale $P_{jm}$ to
$$(-1)^{j+m} { 1\over (2j+1)!} \left[ { (2j)!\over (j+m)!(j-m)!}\right]^{1/2}
P_{jm} , \eqn\xab$$
and $Y^-_{jm}$ to
$$(-1)^{j+m} {1\over \sqrt{j/2} } { 1\over (2j-1)! }
\left[ { (2j-1)!\over (j+m)!(j-m)! } \right]^{1/2} Y^-_{jm} .\eqn\xac$$

Finally let us calculate the product $O_{j_1m_1} P_{j_2m_2}$. In the
course of this calculation we will find a proof of $\newf$.
Generally we should have
$$\eqalign{
O_{j_1m_1}P_{j_2m_2} =& g_1(j_1, j_2)
\langle j_1m_1j_2m_2\mid j_2-j_1,m_1+m_2\rangle
P_{j_2-j_1, m_1+m_2} \cr
& + g_2(j_1, j_2) \langle j_1m_1j_2m_2\mid j_2-j_1+1, m_1+m_2\rangle
aY^-_{j_2-j_1+1, m_1+m_2}  .
} \eqn\xad$$
We will compute $g_1(j_1,j_2)$ only, which can be deduced from
$g_1(1/2, j_2)$. The other unknown function can be derived by using
associativity. For $j_1=-m_1=1/2$ and $m_2=j_2=j$ we have
$$ \eqalign{
O_{1/2, -1/2} P_{j,j}(\delta) =&
\left\{ (2j+\delta) \sum_{q=1}^{2j} { \partial^{q+1} c \over (q+1)! }
 c S_{2j-q} (-iX^+-i\delta X^-) \right.\cr
 &\left.  + \sum_{q=0}^{2j} { \partial^{q+1} c \over (q+1)! } c S_{2j-q}
(-iX^+-i\delta X^-)  \right\}
e^{-2iX^++ i(2j-1)X^- } \cr
=& { (2j+1)(\delta + 2j) \over (2j+1) } P_{j-1/2, j-1/2} (\delta) + \cdots ,
} \eqn\xae$$
where $\cdots$ denotes the term proportional to $aY^-_{j+1/2, j-1/2}$.
Setting $P_{j,j}(\delta) =F_j(\delta)P_{j,j}$, one knows that $F_j(\delta)$
is a polynomial function with the highest degree $2j$ and $F_j(1)$ is 1. From
$\xae$ we get
$$\eqalign{
F_j(\delta) & = { \delta + 2j \over 2j+ 1}F_{j-1/2}(\delta) \cr
& = { \delta + 2j\over 2j+1 } { \delta + 2j-1 \over 2j } \cdots
{\delta+1 \over 2} F_0(\delta) = { (\delta +2j)!\over (2j+1)! \delta !} .
} \eqn\xaf$$
This proves $\newf$. Substituting this result back into $\xae$ one has
$$ O_{1/2, -1/2} P_{j,j} = (2j+2)P_{j-1/2, j-1/2} + \cdots . \eqn\xag$$
For $O_{j_1,-j_1} = -(-1)^{2j_1} (O_{1/2, -1/2})^{2j_1}/(2j_1)^2$,
we have then
$$ O_{j_1, -j_1} P_{j_2, j_2} = -(-1)^{2j_1} { (2(j_2+1))! \over (2j_1)!
(2(j_2-j_1+1))! } P_{j_2-j_1, j_2-j_1} + \cdots . \eqn\xah$$
 From this result it follows
$$\eqalign{
O_{j_1m_1}P_{j_2m_2} = & { j_2+1\over j_2-j_1 + 1} P_{j_2-j_1, m_1+m_2} \cr
& + { \tilde{ g} _2({j_1,j_2}) 1\over j_2-j_1 + 1}
(m_1j_2+m_2j_1+m_1) aY^-_{j_2-j_1+1, m_1+m_2} ,
} \eqn\xai$$
after rescaling $O_{j_1,m_1}$ and $P_{j_2,m_2}$ as in $\okog$ and $\xab$.
The unknown function $\tilde{g}_2(j_1, j_2)$ will be determined in a moment.

Let us summarize the ring structures we have determined. First we have
$$\eqalign{
O_{j_1m_1} O_{j_2m_2} = & O_{j_1+j_2, m_1+m_2} ,  \cr
O_{j_1m_1}Y^+_{j_2m_2}= & {j_2\over j_1+j_2} Y^+_{j_1+j_2, m_1+m_2}
- { 1\over j_1+j_2} (m_1j_2-m_2j_1)aO_{j_1+j_2-1, m_1+m_2} , \cr
Y^+_{j_1m_1}Y^+_{j_2m_2} =& {1\over j_1+j_2-1} (m_1j_2-m_2j_1)
 aY^+_{j_1+j_2-1, m_1+m_2},  \cr
Y^+_{j_1m_1}Y^-_{j_2m_2} = & {j_1\over j_2-j_1+1} P_{j_2-j_1, m_1+m_2} \cr
& + { 1\over j_2-j_1+1} (m_1j_2+m_2j_1+m_1) aY^-_{j_2-j_1+1, m_1+m_2}.
}\eqn\xax$$
The other non-vanishing  products are
$$ \eqalign{
O_{j_1m_1} Y^-_{j_2m_2} = & g_1(j_1, j_2) Y^-_{j_2-j_1, m_1+m_2} , \cr
Y^+_{j_1m_1} P_{j_2m_2} = & { g_2(j_1, j_2) \over j_2-j_1+2}
(m_1j_2+m_2j_1+m_1) aP_{j_2-j_1 +1, m_1+m_2} , \cr
O_{j_1m_1}P_{j_2m_2} = & { j_2+1\over j_2-j_1 + 1} P_{j_2-j_1, m_1+m_2} \cr
& + { \tilde{ g}_2({j_1,j_2}) \over j_2-j_1 + 1} (m_1j_2+m_2j_1+m_1)
aY^-_{j_2-j_1+1, m_1+m_2} ,
} \eqn\xaj$$
with
$$g_1(j_1,j_2)=g_2(j_1,j_2)=\tilde{g}_2(j_1,j_2)=1. \eqn\wuadd$$
while the vanishing products are
$$ \eqalign{
Y^-_{j_1m_1} Y^-_{j_2m_2} =&  0 , \cr
Y^-_{j_1m_1}P_{j_2m_2}  = & 0, \cr
P_{j_1m_1} P_{j_2m_2} = & 0.
} \eqn\xak$$

The three unknown functions in $\xaj$ are determined from associativity.
Multiplying both sides of $\poaa$ from the right by $Y^-_{j_3m_3}$
and using the known result $\xaa$ and the second equation of $\xaj$, we get
$g_2(j_1,j_3-j_2)=1$ or $g_2(j_1, j_2)=1$. To determine
$\tilde{g}_2(j_1, j_2)$ we multiply  the third equation of $\xaj$ by
$Y^+_{j_3m_3}$ and using $\xxx$ and the second equation of $\xaj$. The
result is $\tilde{g}_2(j_1, j_2)= 1$. Finally $g_1(j_1, j_2)$ is
determined by multiplying the first equation of $\xaj$  by
$aY^+_{j_3m_3}$, resulting in $g_1(j_1, j_2)=1$. This result is also
confirmed by explicit calculation.

\def\cgj{\langle j_1m_1 j_2m_2 |j_1+j_2,m_1+m_2 \rangle }
\def\cgjp{\langle j_1m_1 j_2m_2 |j_1+j_2-1,m_1+m_2 \rangle }
\def\cgpj{\langle j_1m_1 j_2m_2 |j_2-j_1,m_1+m_2 \rangle }
\def\cgpjp{\langle j_1m_1 j_2m_2 |j_2-j_1+1,m_1+m_2 \rangle }

To conclude this section let us note that we have chosen an $m$ dependent
normalization for all the vertex operator $O_{jm}$, $Y^{\pm}_{jm}$ and
$P_{jm}$'s. Here we also give the structure equations in their $SU(2)$
covariant form, which are obtained from $\xax$ and $\xaj$ by the following
replacement:
$$\eqalign{
O_{jm}\quad & \to \quad C(jm)O_{jm}, \cr
Y^+_{jm}\quad & \to \quad N(jm) Y^+_{jm},  \cr
Y^-_{jm}\quad & \to \quad
(-1)^{j+m} {\sqrt{j(2j+1)} \over N(jm) } Y^-_{jm}, \cr
P_{jm}\quad & \to \quad
(-1)^{j+m} {\sqrt{j(2j+1)} \over C(jm) } P_{jm}, } \eqn\nokl$$
where $C(jm) = \left[ { (j+m)!(j-m)!\over (2j)! } \right]^{1/2}$ and
$N(jm) = \left[ { (j+m)!(j-m)!\over (2j-1)! } \right]^{1/2}$.
We have then
$$\eqalign{
O_{j_1m_1} O_{j_2m_2} = & \cgj O_{j_1+j_2, m_1+m_2} ,  \cr
O_{j_1m_1}Y^+_{j_2m_2}= & \sqrt{ j_2\over j_1+j_2} \cgj
 Y^+_{j_1+j_2, m_1+m_2} \cr
& -\sqrt{j_1\over j_1+j_2} \cgjp aO_{j_1+j_2-1, m_1+m_2} , \cr
Y^+_{j_1m_1}Y^+_{j_2m_2} =& \sqrt{j_1+j_2\over j_1+j_2-1} \cgjp
 aY^+_{j_1+j_2-1, m_1+m_2},  \cr
Y^+_{j_1m_1}Y^-_{j_2m_2} = & \sqrt{ j_1\over j_2-j_1+1}\cgpj
P_{j_2-j_1, m_1+m_2} \cr
& -\sqrt{ j_2+1\over j_2-j_1+1} \cgpjp aY^-_{j_2-j_1+1, m_1+m_2},  \cr
O_{j_1m_1} Y^-_{j_2m_2} = & \cgpj Y^-_{j_2-j_1, m_1+m_2} , \cr
Y^+_{j_1m_1} P_{j_2m_2} = & -\sqrt{ j_2-j_1+1\over j_2-j_1+2 } \cgpjp
aP_{j_2-j_1 +1, m_1+m_2} , \cr
O_{j_1m_1}P_{j_2m_2} = & \sqrt{j_2+1\over j_2-j_1+1} \cgpj
 P_{j_2-j_1, m_1+m_2} \cr
& - \sqrt{ j_1\over j_2-j_1+1} \cgpjp aY^-_{j_2-j_1+1, m_1+m_2}.
} \eqn\xppaj$$
The other three vanishing products remain unchanged.

\vfill\eject

\centerline{ \bf 4. Chiral Ring Structure Including the Tachyon States}

In this section we incorporate the tachyon states into the chiral
cohomology ring. The tachyon states are labelled by a continuous momentum
$p$ and there are two chiralities:
$$ T^{ (\pm) }_p = ce ^{ ipX+ (\sqrt{2} \mp p) \phi } . \eqn\qaa$$
As in the case of discrete states, there also exist absolute tachyon states
$$ aT^{ (\pm) }_p = \pm { p \over \sqrt{2} } \partial c
ce ^{ ipX+ (\sqrt{2} \mp p) \phi } . \eqn\qab$$
To simplify our presentation we will consider only those tachyon states
which are not included in the discrete states, i.e. $p\neq j\sqrt{2}$, where
$j$ is an integer or half integer. Let us first consider the product of
discrete states with tachyon states. By momentum addition this product must
give rise to another tachyon state. Since any tachyon states must be of
the form given by either $\qaa$ or $\qab$, there are only four
non-vanishing products of this type:
$$ \eqalign{
O_{j, \pm j} T^{(\pm)}_p  & \sim  T^{(\pm )} _{p\pm j\sqrt{2}} , \cr
Y^+_{j, \pm (j-1) } T^{(\pm)}_p  & \sim  a  T^{(\pm )}_{p\pm (j-1)\sqrt{2}} .
} \eqn\qac$$
The explicit calculation of these products is easy and we get, in
agreement with [\WITZ],
$$ \eqalign{
O_{j,  j} T^{(+)}_p  =& { p \over p + \sqrt{2}j }
T^{(+ )}_{p+ j\sqrt{2}} , \cr
O_{j, -j} T^{(-)}_p = & T^{(-)}_{p-\sqrt{2} j} , \cr
Y^+_{j , (j-1) } T^{(+)}_p  =&  - { p\over p +\sqrt{2}(j-1) }
 a  T^{(+ )}_{p+\sqrt{2}(j-1)} , \cr
Y^+_{j , -(j-1) } T^{(-)}_p  =&  -
 a  T^{(- )}_{p- \sqrt{2}(j-1)} , \cr
 } \eqn\qad$$
if we rescale $T^{(\pm)}$ to
$$  (\Gamma(1+\sqrt{2}p) ) ^{\pm 1}  T^{ (\pm)} _p . \eqn\pae$$

Now let us consider the product of two tachyons. For the product
$T^{(-)}_p T^{(-)}_q$ the exponential factor is $e^{ i(p+q)X + (2\sqrt{2} +
(p+q))\phi} $ which can not be the exponential factor of any tachyon state. To
be able to obtain a discrete state we must have
$$ p+ q = j \sqrt{2}, \eqn\qaf$$
where $j $ is an integer or half integer. For $j\ge -1/2$ we have
$$  T^{(-)}_p T^{(-)}_q = d_1aY^-_{j+1, j} + d_2P_{j, j} ,  \eqn\qag$$
where $d_1$ and $d_2$ are two constants.
For $j \le -1$ the product is zero, for there is no $aY^+_{-j -1, j}$. By
explicit calculation we have
$$ \eqalign{
T^{(-)}_pT^{(-)}_q = & \sum_{k=0} ^{2j+1} { \partial^{k+1} c \over (k+1)! } c
S_{2j+1-k } (-iX^+ +i(\sqrt{2} p +1) X^- ) e^{ (ijX+(2+j)\phi)\sqrt{2} } \cr
=& { \Gamma(2j+1-\sqrt{2} p) \over 2(j+1) \Gamma(-\sqrt{2} p) }
\Big(P_{j, j} + aY^-_{j +1, j} \Big),
} \eqn\qah$$
or
$$
T^{(-)}_p T^{(-)}_q = { \sin \sqrt{2}\pi p \over 2(j+1) \pi }
\Big(P_{j, j} + aY^-_{j +1, j} \Big),
\eqn\qai$$
after making rescaling $\pae$. Similarly we have, with $p+q= -j\sqrt{2}$,
$$ T^{(+)}_pT^{(+)}_q ={  \pi pq \over (j+1) \sin\sqrt{2}\pi q}
\Big(P_{j, -j} - aY^-_{j +1, -j} \Big).
\eqn\qaj$$

The product $T^{(+)}_p T^{(-)}_q$ gives no new result. It either reduces
to $\qad$ or to $Y^+_{j_1, \pm j_1} Y^-_{j_2, \pm j_2}$,
which is a special case of $\xaa$.

Finally let us point out that part of the above results was also
obtained in [\TAN]. There they computed the single pole term in the OPE of
$\Psi^{(\pm)}_p \Psi^{(\pm)}_q$, where $T_p ^{(\pm)} = c\Psi^{(\pm)}_p$.
This only gives the term $aY^-_{j+1, \pm j} $ in $\qai$ or $\qaj$.
The other term $P_{j, \pm j} $ comes from the higher order
pole terms in the OPE of $\Psi^{(\pm)}_p \Psi^{(\pm)}_q$.

\vfill\eject

\centerline{ \bf 5. Chiral Transformation Rules and the Chiral
Charge  Algebra}

In this section we will use the above results to study the (chiral)
symmetry charge algebra associated to the chiral states. We will follow
the general strategy of constructing BRST invariant charges
from BRST invariant operators [\WITZ]. We consider only the holomorphic
part in this section. In the next section we will use the results to
derive the transformation rules for the discrete closed-string states by
combining the left and right sectors.

Starting from a generic BRST invariant operator $V_{jm}$, one defines
$V^{(1)}_{jm}$ by the OPE
$$ b(z)V_{jm}(w) \sim \cdots + { 1\over z-w} V_{jm}^{(1)} , \eqn\raa$$
or
$$ V^{(1)}_{jm}(w) = \oint_0 [\dd z] b(z)V_{jm}(w),   \eqn\rab$$
up to total derivatives and BRST exact terms. Then one can prove
$$ \partial_w V_{jm}  (w) = \{ Q, V^{(1)}_{jm} (w) \} . \eqn\rac$$
Indeed,
$$\eqalign{
\{ Q, V^{(1)}_{jm}& (w) \} =  \oint_w [\dd z]
:c(z) \big( T(z) +\partial_z c(z)
b(z) \big):  \oint_w [\dd \tilde{z} ] b(\tilde{z})V_{jm}(w) \cr
= & \Big( \oint_w[\dd \tilde{z}]\oint_w [\dd z] + \oint_w[\dd \tilde{z}]
\oint_{\tilde{z}} [\dd z] \Big)
:c(z) \big( T(z) +\partial_z c(z)
b(z) \big):  b(\tilde{z})V_{jm}(w) \cr
= & \oint_w[\dd z] \Big( T(z) + :2\partial_z c(z)b(z)+ c(z)\partial_zb(z)
:\Big) V_{jm}(w) \cr
= & \partial_w V_{jm}(w) .
}\eqn\rad$$
In terms of $V^{(1)}_{jm}$ one can define a BRST invariant charge
$${ \hat{V}} _{jm} = \oint_0[\dd z] V^{(1)}_{jm} (z). \eqn\rae$$
The (anti-)commutator of $\hat{V}_{jm}$ with any BRST invariant local operator
gives another BRST invariant operator, i.e.
$$ [\hat{V}_{j_1m_1} , V_{j_2m_2}(z) ] = \sum_{jm} g_(j_1, j_2; j)
\langle j_1m_1j_2m_2\mid jm\rangle V_{jm}(z) . \eqn\raf$$

Let us see how one can compute the RHS of $\raf$ explicitly from
the knowledge of the cohomology ring structure. Let us consider first
the relative physical states. For them the double pole term in $\raa$
actually vanishes, because there is no $\partial c$ term in $V_{jm}$.
By definition, the ring multiplication law is given by
$$
(V_{j_1m_1}V_{j_2m_2}) (w)
= \oint_w[\dd z] { 1\over z-w} V_{j_1m_1} (z) V_{j_2m_2} (w).
\eqn\rag$$
Applying $\oint_w[\dd \tilde{z}] (\tilde{z} -w) b(\tilde{z}) $
on both sides from the left, we get
$$ \eqalign{
{\rm RHS} = &  \Big( \oint_w[\dd z] \oint_w[\dd \tilde{z} ] +
\oint_w[\dd z] \oint_z[\dd \tilde{z} ]\Big) { \tilde{z} -w \over z-w}
b(\tilde{z} ) V_{j_1m_1}(z) V_{j_2m_2}(w) \cr
= & \oint_w[\dd z] V^{(1)}_{j_1m_1} (z) V_{j_2m_2}(w) = [\hat{V}_{j_1m_1},
V_{j_2m_2}(w) ] ;
} \eqn\rah$$
Thus we have derived the important formula
$$ [\hat{V}_{j_1m_1} , V_{j_2m_2} ] = \oint_w [\dd z] (z-w)b(z)(V_{j_1m_1}
V_{j_2m_2})(w). \eqn\rahh$$
This formula allows us to compute the action of a charge on a physical
states from the knowledge of the cohomology ring structure, exhibited by
$$
(V_{j_1m_1}V_{j_2m_2}) (w)
= \sum_{jm } \tilde{g}(j_1, j_2; j)\langle j_1m_1j_2m_2\mid jm\rangle
V_{jm}(w). \eqn\wuadd $$
In fact, by using $\wuadd$ the RHS of $\rahh$ is simply given by
$$ \hbox{ RHS of}~~ \rahh = \sum_{jm} \tilde{g}(j_1, j_2; j)
\langle j_1m_1 j_2m_2\mid
jm\rangle \oint_w[\dd z] (z-w)b(z) V_{jm}(w). \eqn\rai$$
We note that the last integration selects only terms with a factor
$\partial c$ in $V_{jm}$. In this way we see that the transformation
rules for the discrete charges acting on the discrete states can be
simply read off from the structure equation (i.e. the multiplication law)
of the cohomology ring.

Now let us apply the above results to the discrete states
$O_{jm}$, $Y^{\pm}_{jm}$ and $P_{jm}$. From these discrete states
one can construct BRST invariant charges of ghost number $-1$, $0$
and $1$. We denote these charges as
$\hat{X}_{jm}$, $Q^{\pm}_{jm}$ and $\hat{Z}_{jm}$:
$$ \eqalign{
\hat{X}_{jm} \equiv &  \oint_0[\dd w] \oint_w[\dd z] b(z) O_{jm}(w)
\equiv \oint_0 [\dd w] X_{jm}(w), \cr
Q^{\pm}_{jm} \equiv &  \oint_0[\dd w] \oint_w[\dd z] b(z) Y^{\pm}_{jm}(w)
\equiv \oint_0 [\dd w] W^{\pm}_{jm}(w), \cr
\hat{Z}_{jm} \equiv &  \oint_0[\dd w] \oint_w[\dd z] b(z) P_{jm}(w)
\equiv \oint_0 [\dd w] Z_{jm}(w).} \eqn\epxd$$
Their action on discrete states can be directly read off from the
ring structure by using $\rahh$. We have
$$ \eqalign{
& [\hat{X}_{j_1m_1} , O_{j_2m_2} ] = 0, \qquad
\{ \hat{X}_{j_1m_1} , Y^-_{j_2m_2}\} = 0, \cr
& \{ \hat{X}_{j_1m_1} , Y^+_{j_2m_2} \}
= -(m_1j_2- m_2j_1)O_{j_1+j_2-1, m_1+m_2}, \cr
& [\hat{X}_{j_1m_1}, P_{j_2m_2}]
= -(m_1j_2+m_2j_1+m_1)Y^-_{j_2-j_1+1, m_1+m_2}, \cr
& [Q^+_{j_1m_1}, Y^+_{j_2m_2} ]
= (m_1j_2-m_2j_1)Y^+_{j_1+j_2-1, m_1+m_2}, \cr
& [Q^+_{j_1m_1}, Y^-_{j_2m_2} ]
= -(m_1j_2+m_2j_1+m_1)Y^-_{j_2-j_1+1, m_1+m_2} , \cr
& [Q^+_{j_1m_1}, P_{j_2m_2}]
= -(m_1j_2+m_2j_1+m_1)P_{j_2-j_1+1, m_1+m_2}, \cr
& \{ Q^-_{j_1m_1}, Y^-_{j_2m_2} \} = 0, \quad
[Q^-_{j_1m_1}, P_{j_2m_2} ] = 0, \quad
[ \hat{Z}_{j_1m_1} , P_{j_2m_2} ] =0 .
}\eqn\raj$$
Applying once more
$\oint_0[\dd w] \oint_w[\dd z] b(z)$ from the
left on these equations\footnote*{This operation changes all the discrete
states into the corresponding charges.}, we get the following chiral charge
algebra
$$ \eqalign{
& [ \hat{X}_{j_1m_1} , Q^+_{j_2m_2}]
= (m_1j_2- m_2j_1)\hat{X}_{j_1+j_2-1, m_1+m_2}, \cr
& \{\hat{X}_{j_1m_1}, \hat{Z}_{j_2m_2}\}
= (m_1j_2+m_2j_1+m_1)Q^-_{j_2-j_1+1, m_1+m_2}, \cr
& [Q^+_{j_1m_1}, Q^+_{j_2m_2} ]
= (m_1j_2-m_2j_1)Q^+_{j_1+j_2-1, m_1+m_2}, \cr
& [Q^+_{j_1m_1}, Q^-_{j_2m_2} ]
= -(m_1j_2+m_2j_1+m_1)Q^-_{j_2-j_1+1, m_1+m_2} , \cr
& [Q^+_{j_1m_1}, \hat{Z}_{j_2m_2}]
= -(m_1j_2+m_2j_1+m_1)\hat{Z}_{j_2-j_1+1, m_1+m_2}. \cr
}\eqn\rak$$
The remaining (anti-)commutators are zero.

Similarly one can also construct BRST invariant charges from absolute physical
states. We simply denote these absolute charges as $a\hat{X}_{jm}$,
$aQ^{\pm}_{jm}$ and
$a\hat{Z}_{jm}$\footnote{**}{$a\hat{X}_{jm}\equiv \oint_0[\dd w]\oint_w[\dd z]
b(z)(aO_{jm})(w)\neq a\cdot \hat{X}_{jm},$ etc. }. Then one easily proves the
following general relations
$$\eqalign{
[\hat{V}_{j_1m_1}, aV_{j_2m_2}(w)] = & -(-1)^{[V_{j_1}]} g_V(j_2)
(V_{j_1m_1}V_{j_2m_2})(w) \cr
& + \oint_w[\dd z](z-w)b(z)(V_{j_1m_1}aV_{j_2m_2})(w) , \cr
[a\hat{V}_{j_1m_1}, V_{j_2m_2}(w)] = & - g_V(j_1)
(V_{j_1m_1}V_{j_2m_2})(w) \cr
& + \oint_w[\dd z](z-w)b(z)(aV_{j_1m_1}V_{j_2m_2})(w) , \cr
[a\hat{V}_{j_1m_1}, aV_{j_2m_2}(w)] = & -(-1)^{[V_{j_1}]}
(g_V(j_1)- g_V(j_2) (aV_{j_1m_1}V_{j_2m_2})(w) , \cr } \eqn\zza$$
just by deforming the integration contour as in $\rah$.  Here in the first
and third equations $[V_{j_1}]$ denotes the ghost number of $V_{j_1m_1}$.
The function $g_V(j_a)$ ($a=1,2$) is defined as
$$ g_O(j_a) = -  g_P(j_a) = j_a+1, \qquad g_{Y^{\pm}} (j_a) = \pm j_a ,
 \eqn\zzb$$
depending on the type of the operator $V_{j_a m_a}$.

Now we present some examples to see how the above formulas can be
used to derive the chiral transformation rules and chiral charge
algebra involving absolute physical states or absolute charges. We have,
e.g.,
$$ \eqalign{
[\hat{X}_{j_1m_1} , aY^+_{j_2m_2} ]  = &
{j_1j_2\over j_1 + j_2} Y^+_{j_1+j_2, m_1+m_2}
+ {j_2 \over j_1+j_2}(m_1j_2-m_2j_1) aO_{j_1+j_2-1, m_1+m_2} , \cr
[aQ^+_{j_1m_1}, O_{j_2m_2} ] = &
{j_1j_2\over j_1 + j_2} Y^+_{j_1+j_2, m_1+m_2}
- {j_1 \over j_1+j_2}(m_1j_2-m_2j_1) aO_{j_1+j_2-1, m_1+m_2} , \cr
[Q^+_{j_1m_1}, aO_{j_2m_2}] = &
-{j_1(j_1-1) \over j_1+j_2} Y^+_{j_1+j_2, m_1+m_2}
\cr
& + {j_2+1\over j_1+j_2} (m_1j_2 -m_2j_1)aO_{j_1+j_2-1, m_1+m_2} , \cr
[a\hat{X}_{j_1m_1}, Y^+_{j_2m_2} ] = &
{j_2(j_2-1) \over j_1+j_2} Y^+_{j_1+j_2, m_1+m_2}
\cr
& + {j_1+1\over j_1+j_2} (m_1j_2 -m_2j_1)aO_{j_1+j_2-1, m_1+m_2}
. } \eqn\zzc$$
Applying $\oint_0[\dd w] \oint_w [\dd z] b(z)$ from the
left to the above equations, we get the following (chiral)
(anti-)commutation relations
$$ \eqalign{
\{\hat{X}_{j_1m_1} , aQ^+_{j_2m_2} \} = &
-{j_1j_2 \over j_1 + j_2} Q^+_{j_1+j_2, m_1+m_2}
\cr
& + {j_1\over j_1 +j-2}(m_1j_2 - m_2j_1)a\hat{X}_{j_1+j_2 -1, m_1+m_2} , \cr
[Q^+_{j_1m_1}, a\hat{X}_{j_2m_2}] = &
-{j_1(j_1-1)\over j_1+j_2} Q^+_{j_1+j_2, m_1+m_2}
\cr
& + {j_2\over j_1 +j-2}(m_1j_2 - m_2j_1)a\hat{X}_{j_1+j_2 -1, m_1+m_2} . }
\eqn\zzd$$

We have derived the complete set of the chiral transformation rules
and the chiral charge algebra and collect them in appendix B.

\vfill\eject

\centerline{ \bf 6. Symmetry Transformation of the Discrete String States}

All of our discussions up to now deal with the right-moving sector only.
In this section we will combine the left and right sectors to derive
the symmetry transformation rules for the discrete states in closed string
theory. Before doing this let us recall briefly the physical states
in $D=2$ closed string theory.

As discussed in [\WITZ], the physical closed string states in $D=2$
should satisfy an extra condition
$(b_0-\bar{b}_0 ) |\hbox{phys.} \rangle = 0$.
We will not go into any details for solving this condition
but simply quote their results. By matching the Liouville momenta,
the physical closed string discrete states that obey
the $b_0-\bar{b}_0$ condition are as follows ($G$ is the total
ghost number):

\noindent 1) the plus states
$$\eqalign{
& G=0: \quad O_{jm}\bar{O}_{jm'} ;\cr
& G=1: \quad O_{jm}\bar{Y}^+_{j+1, m'},\quad Y^+_{j+1, m}\bar{O}_{jm'} ,
\quad (a+ \bar{a}) O_{jm}\bar{O}_{jm'} ;\cr
& G=2: \quad Y^+_{jm}\bar{Y}^+_{jm'},
\quad (a+\bar{a}) O_{jm}\bar{Y}^+_{j+1, m'},
\quad (a+\bar{a})Y^+_{j+1, m}\bar{O}_{jm'} ; \cr
& G=3: \quad (a+ \bar{a})Y^+_{jm}\bar{Y}^+_{jm'}; } \eqn\zze$$

\noindent 2) the minus states
$$\eqalign{
& G=2: \quad Y^- _{jm}\bar{Y}^-_{jm'} ;\cr
& G=3: \quad P_{jm}\bar{Y}^-_{j+1, m'},\quad Y^-_{j+1, m}\bar{P}_{jm'} ,
\quad (a+ \bar{a}) Y^-_{jm}\bar{Y}^-_{jm'} ;\cr
& G=4: \quad P_{jm}\bar{P}_{jm'}, \quad (a+\bar{a}) P_{jm}\bar{Y}^-_{j+1, m'},
\quad (a+\bar{a})Y^-_{j+1, m}\bar{P}_{jm'} ; \cr
& G=5: \quad (a+ \bar{a})P_{jm}\bar{P}_{jm'}. } \eqn\zzf$$
 From the above list we see that half of the states (the states without $(a+
\bar{a})$) are just the products of left and right relative physical states
which satisfy the stronger condition:
$b_0|\hbox{phys.}\rangle=\bar{b}_0|\hbox{phys.}\rangle=0$. For later
convenience we call these states relative physical (string) states.
The other half of the states contain $(a +\bar{a})$ and do not
satisfy either $b_0|\hbox{phys.}\rangle=0$ or
$\bar{b}_0|\hbox{phys.}\rangle=0$. We will call them absolute
physical (string) states. For closed string states, all the absolute
physical states can be obtained from the relative physical states
simply by multiplying $(a+\bar{a})$.

The general framework for constructing conserved charges in $D=2$
closed string theory again has been given in sec. 4 of [\WITZ].
Summarizing their results, one starts with any BRST invariant zero
form $\Omega^{(0)}$ and constructs the one and two forms $\Omega^{(1)}$
and $\Omega^{(2)}$ satisfying the descent equations
$$ \eqalign{
0 = & \{ Q_T,  \Omega^{(0) } \} , \cr
\dd\Omega^{(0)} = & \{ Q_T, \Omega^{(1)} \} , \cr
\dd\Omega^{(1)} = & \{ Q_T, \Omega^{(2)} \} . } \eqn\zzg$$
These equations imply that the charge
$ A = { 1\over 2\pi i} \oint_0 \Omega^{(1)} $
is a BRST invariant charge (the second equation)
conserved up to BRST trivial operators (the third equation). $\Omega^{(1)}$
and  $\Omega^{(2)}$ are given by\footnote*{ $\oint_w[\dd z] { 1\over
z-w} = -\oint_w[\dd \bar{z}] { 1\over \bar{z}-\bar{w} } = 1.$ }
$$\eqalign{
 \Omega^{(1)}(w, \bar{w}) = & \Big(\dd w \oint_w [\dd z] b(z) - \dd \bar{w}
\oint_w[\dd \bar{z}] \bar{b}(\bar{z})\Big)  \Omega^{(0)}(w, \bar{w}), \cr
 \Omega^{(2)}(w, \bar{w}) = & - \dd w\wedge \dd \bar{w} \oint_w[\dd z']
b(z') \oint_w[\dd \bar{z} ] \bar{b} (\bar{z}) \Omega^{(0)} (w,\bar{w}) . }
\eqn\zzh$$

In what follows we take $\Omega^{(0)}$ to be $Y^+_{j+1, m}\bar{O}_{jm'}$
as an example and derive the symmetry transformations for discrete string
states. We have then
$$ \eqalign{
  \Omega^{(1)}(w, \bar{w}) = & \dd w W^+_{j+1, m } \bar{O}_{jm'}(w, \bar{w})
- \dd \bar{w} Y^+_{j+1, m} \bar{X}_{jm'}(w, \bar{w}), \cr
  \Omega^{(2)}(w, \bar{w}) = & - \dd w\wedge \dd \bar{w} W^+_{j+1, m}
\bar{X}_{jm'} (w, \bar{w}) . } \eqn\zzi$$
The conserved charge $A_{j; mm'}\equiv{ 1\over 2 \pi i }
\oint_0  \Omega^{(1)}(w, \bar{w}) $ acting on the (plus) relative discrete
states gives rise to
$$\eqalign{
[A_{j_1;m_1m_1'},& O_{j_2m_2}\bar{O}_{j_2m_2'}] \cr
=  & (m_1j_2-m_2(j_1+1))O_{j_1+j_2,m_1+m_2} \bar{O}_{j_1+j_2, m_1'+m_2'}, \cr
[A_{j_1;m_1m_1'},& O_{j_2m_2}\bar{Y}^+_{j_2+1, m_2'}]\cr
= & -{1\over j_1+j_2+1}(m_1j_2-m_2(j_1+1))(m_1'(j_2+1)-m_2'j_1) \cr &\times
(a +\bar{a} )O_{j_1+j_2, m_1+m_2}\bar{O}_{j_1+j_2, m_1'+m_2'}  \cr
& + {j_2 + 1\over j_1+j_2 + 1 }(m_1j_2 -m_2(j_1+1))O_{j_1+j_2, m_1+m_2}
\bar{Y}^+_{j_1+j_2+1, m_1'+m_2'}\cr
&-{ j_1+1\over j_1+j_2+1} (m_1'(j_2+1)-m_2'j_1)Y^+_{j_1+j_2+1,m_1+m_2}
\bar{O}_{j_1+j_2, m_1'+m_2'}, \cr
[A_{j_1;m_1m_1'},& Y^+_{j_2+1,m_2}\bar{O}_{j_2m_2'}] \cr = &
(m_1(j_2+1)-m_2(j_1+1)) Y^+_{j_1+j_2+1, m_1+m_2}
\bar{O}_{j_1+j_2, m_1'+m_2'},\cr
[A_{j_1;m_1m_1'},& Y^+_{j_2m_2}\bar{Y}^+_{j_2m_2'} ] \cr = &
{1\over j_1+j_2} (m_1j_2-m_2(j_1+1))  (m_1'j_2-m_2'j_1)\cr &\times
(a +\bar{a}) Y^+_{j_1+j_2, m_1+m_2}\bar{O}_{j_1+j_2-1, m_1'+m_2'}\cr
& + { j_2\over j_1+j_2} (m_1j_2-m_2(j_1+1)) Y^+_{j_1+j_2,m_1+m_2}\bar{Y}^+_
{j_1+j_2, m_1'+m_2'} .}
\eqn\zzp$$
Note that the action of the charge $A_{j_1; m_1m_1'}$ on the relative
discrete states also give rise to absolute physical states.
This feature was noted by Witten and Zwiebach [\WITZ] too.
The example given by them is a special case of the last
equation in $\zzp$. Here we are able to systematically
derive the complete set of transformation rules with the help of the
knowledge of the cohomology ring structure. In appendix C we will show that
the action of any charge on the physical closed string discrete states gives
only (a linear combination of) physical closed string discrete states,
i.e., those listed in $\zze$ and $\zzf$.

The action of the charge $A_{j_1; m_1m_1'}$ on the (plus) absolute discrete
states can also be derived. We have
$$\eqalign{
[A_{j_1;m_1m_1'},& (a+\bar{a})O_{j_2m_2}\bar{O}_{j_2m_2'}]\cr = &
(m_1j_2-m_2(j_1+1))(a+\bar{a})
O_{j_1+j_2,m_1+m_2} \bar{O}_{j_1+j_2, m_1'+m_2'}, \cr
[A_{j_1;m_1m_1'},& (a+ \bar{a}) O_{j_2m_2}\bar{Y}^+_{j_2+1, m_2'}] \cr =
&-{ j_1+1\over j_1+j_2+1} (m_1'(j_2+1)-m_2'j_1)
(a +\bar{a}) Y^+_{j_1+j_2+1,m_1+m_2}
\bar{O}_{j_1+j_2, m_1'+m_2'} \cr
& + {j_2 + 1\over j_1+j_2 + 1 }(m_1j_2 -m_2(j_1+1))
(a+\bar{a}) O_{j_1+j_2, m_1+m_2}
\bar{Y}^+_{j_1+j_2+1, m_1'+m_2'}, \cr
[A_{j_1;m_1m_1'},& (a+\bar{a})Y^+_{j_2+1,m_2}\bar{O}_{j_2m_2'}] \cr = &
(m_1(j_2+1)-m_2(j_1+1)) (a+\bar{a}) Y^+_{j_1+j_2+1, m_1+m_2}
\bar{O}_{j_1+j_2, m_1'+m_2'},\cr
[A_{j_1;m_1m_1'},& (a+\bar{a}) Y^+_{j_2m_2}\bar{Y}^+_{j_2m_2'} ] \cr = &
{ j_2\over j_1+j_2} (m_1j_2-m_2(j_1+1)) (a +\bar{a})
Y^+_{j_1+j_2,m_1+m_2}\bar{Y}^+_ {j_1+j_2, m_1'+m_2'}. \cr } \eqn\zzj$$
Comparing $\zzp$ and $\zzj$, we observe that the left sides of all
the equations in $\zzj$ can be obtained from $\zzp$ simply by
the multiplication of $(a+\bar{a})$. (Note that $(a+\bar{a})^2=0$.)
This suggests the following general formula:
$$ [ \hat{V}_{j_1; m_1m_1'}, (a+\bar{a})V_{j_2;m_2m_2'}] = \pm
(a+\bar{a}) [ \hat{V}_{j_1; m_1m_1'}, V_{j_2;m_2m_2'}] , \eqn\pqq$$
where $\hat{V}_{j_1; m_1m_1'}$ is the charge derived from the relative physical
state $V_{j_1;m_1m_1'}$ and $V_{j_2; m_2m_2'}$ is a relative physical state.
Here one takes the ``$-$'' sign if $ \hat{V}_{j_1; m_1m_1'}$ has odd (total)
ghost number. The general validity of $\pqq$ will be proved in Appendix C.

\def\pxxx{ (a+\bar{a}) }

Quite similarly  one can construct conserved charges from absolute discrete
states and study their action on discrete states.  The corresponding charges
will simply be denoted as $  (a+\bar{a})\hat{V}_{j;mm'}$. Now take
$\Omega^{(0)}$, for example, to be $(a+\bar{a})Y^+_{j+1,m}O_{j,m}$.
By explicit calculation we have
$$\eqalign{
[\pxxx &A_{j_1;m_1m_1'}, O_{j_2m_2}\bar{O}_{j_2m_2'}] \cr
=  &- (m_1j_2-m_2(j_1+1))\pxxx
O_{j_1+j_2,m_1+m_2} \bar{O}_{j_1+j_2, m_1'+m_2'}, \cr
[\pxxx &A_{j_1;m_1m_1'}, O_{j_2m_2}\bar{Y}^+_{j_2+1, m_2'}]\cr
= & -{j_2 + 1\over j_1+j_2 + 1 }(m_1j_2 -m_2(j_1+1))\pxxx O_{j_1+j_2, m_1+m_2}
\bar{Y}^+_{j_1+j_2+1, m_1'+m_2'}\cr
&+{ j_1+1\over j_1+j_2+1} (m_1'(j_2+1)-m_2'j_1)\pxxx Y^+_{j_1+j_2+1,m_1+m_2}
\bar{O}_{j_1+j_2, m_1'+m_2'}, \cr
[\pxxx &A_{j_1;m_1m_1'}, Y^+_{j_2+1,m_2}\bar{O}_{j_2m_2'}] \cr = &
-(m_1(j_2+1)-m_2(j_1+1)) \pxxx Y^+_{j_1+j_2+1, m_1+m_2}
\bar{O}_{j_1+j_2, m_1'+m_2'},\cr
[\pxxx &A_{j_1;m_1m_1'}, Y^+_{j_2m_2}\bar{Y}^+_{j_2m_2'} ] \cr
= & - { j_2\over j_1+j_2} (m_1j_2-m_2(j_1+1))
\pxxx Y^+_{j_1+j_2,m_1+m_2}\bar{Y}^+_
{j_1+j_2, m_1'+m_2'} , }
\eqn\zzp$$
and
$$\eqalign{ & \{ \pxxx A_{j_1;m_1m_1'}, \pxxx O_{j_2m_2}\bar{O}_{j_2m_2'}\} =0,
\cr
& [\pxxx A_{j_1;m_1m_1'}, \pxxx O_{j_2m_2}\bar{Y}^+_{j_2+1, m_2'}]= 0 \cr
& [\pxxx A_{j_1;m_1m_1'}, \pxxx Y^+_{j_2+1,m_2}\bar{O}_{j_2m_2'}]=  0, \cr
& \{ \pxxx A_{j_1;m_1m_1'},
\pxxx Y^+_{j_2m_2}\bar{Y}^+_{j_2m_2'} \} = 0. } \eqn\pxqq$$
 From the above results we would like to guess the following general formulas
$$ \eqalign{& [ \pxxx \hat{V}_{j_1;m_1m_1'} , V_{j_2;m_2m_2'}] = -\pxxx
[ \hat{V}_{j_1;m_1m_1'} , V_{j_2;m_2m_2'}], \cr &
[ \pxxx \hat{V}_{j_1;m_1m_1'} , \pxxx V _{j_2;m_2m_2'}] = 0. } \eqn\pyqq$$
Again the general validity of these equations will be proved in Appendix C.

A remark is in order about the transformation rules in closed string theory.
For the chiral transformation rules we see quite clearly from the general
formulas $\rahh$ and $\zza$ and also the
explicit results in appendix B that the
multiplication by $a$ is not (anti-)commutative with the operation of taking
(anti-) commutator between charge and discrete state,  and also the
operation of getting the charges from the states, i.e. in general
$$\eqalign{ & [\hat{V}_{j_1m_1}, aV_{j_2m_2}] \neq -(-1)^{[V_{j_1}]} a
[\hat{V}_{j_1m_1}, V_{j_2m_2}] , \cr
& [a\hat{V}_{j_1m_1}, V_{j_2m_2}] \neq - a
[\hat{V}_{j_1m_1}, V_{j_2m_2}] , \cr
& [a \hat{V}_{j_1m_1}, aV_{j_2m_2}] \neq 0 .
} \eqn\zhuadda$$
This is in sharp contrast with the transformation rules for closed string
theory. In closed string theory, the $(a+\bar{a})$ plays the role of $a$
played in chiral theory. The eqs. $\pqq$ and $\pyqq$ shows clearly that
the multiplication by $(a+\bar{a})$ can be moved freely and    can be
taken out of the commutator. At the moment we do not have a deep
understanding of this result.

\vglue .5cm

\centerline{ \bf 7. Discussion and Conclusions}

In this paper we have explicitly determined the complete set of
structure constants of the chiral BRST cohomology ring
for $D=2$ string theory. As we demonstrated in sec. 5, this ring
structure encodes  the full chiral charge algebra
and the corresponding symmetry transformation rules. The latter in turn
can be used to derive the full symmetry algebra and transformation rules
for closed string states, as shown in sec. 6. We alos found that the
operator $(a+\bar{a})$ in closed string theory obeys a quite simple rule
as given in eqs. $\pqq$ and $\pyqq$.

\REF\GK{D. Gross and I. Klebanov, \np\ {\bf B352} (1991) 671.}

\REF\MS{G. Moore and N. Seiberg, Rutgers University preprint No.
RU-91-29, 1991.}

\REF\AJ{J. Avan and A. Jevicki, \pl\ {\bf B266} (1991) 35; {\bf B272}
(1991) 17.}

\REF\DDMW{S.R. Das, A. Dhar, G. Mandal and S.R. Wadia, IAS preprint
IASSNS-HEP-91/52.}

\REF\PSR{C. Pope, L. Romans and X. Shen, \pl\ {\bf B236} (1990) 236.}

\REF\WXY{E. Witten, Phys. Rev. {\bf D44} (1991) 314.}

\REF\YWW{F. Yu and Y.-S. Wu, \prl\ {\bf 68} (1992) 2996.}

\REF\BKK{I. Bakas and E. Kiritsis, Maryland/Berkeley/LBL preprint
No. UCB-PTH-91/44, LBL-31213, and UMD-PP-92/37, 1991.}

\REF\YYWW{F. Yu and Y.-S. Wu, \np\ {\bf B373} (1992) 713.}

The search for the infinite symmetries in $D=2$ string theory
has been done recently in various approaches, with apparently
different outcomes. In the $c=1$ matrix model [\GK, \MS, \AJ, \DDMW],
it is the linear $W_{\infty}$ symmetry [\PSR], which is realized through
its well-known free-fermion realization. In the gauged WZW model approach
[\WXY], the first-quantized $D=2$ string theory is described by the
world-sheet $SL(2,R)/U(1)$ model at level $k=9/4$. Recently it has been
shown [\YWW, \BKK, \YYWW] that for generic $k$ there exists a hidden
non-linear $\hat{W}_{\infty}$ [\YWW] current algebra, whose generators
are higher-spin currents formed from the basic coset currents through
an elegant generating function [\YWW]. In this paper we are discussing
the symmetries in the Liouville approach, initiated in [\WITA, \WITZ],
for physical string states. Through we have been able to work out
explicit symmetry transformation rules, the relatioship with the
$W_{\infty}$ symmetries appearing in other approaches as mentioned above
remains to be clarified.

\REF\KAPP{I. R. Klebanov, \mpl\ {\bf A7} (1992) 723.}

\REF\KACH{S. Kachru, \mpl\ {\bf A7} (1992) 1419. }

\REF\PAS{I. R. Klebanov and A. Pasquinucci, ``Correlation Function from
Two Dimensional String Ward Identities,'' preprint PUPT-1313 (April, 1992).}

\REF\BER{M. Bershadsky and D. Kutasov, ``Scattering of Open and Closed
Strings in $1+1$ Dimensions,'' preprint PUPT-1315 or HUTP-92/A016 (April,
1992). }

\REF\FRA{P. Di Francesco and D. Kutasov, \np\ {\bf B375} (1992) 119. }

\REF\CHA{N. Chair, V. K. Dobrev and H. Kanno, \pl\ {\bf B283} (1992) 194.}

\REF\DOTB{Vl. S. Dotsenko, ``Remarks on the Physical States and the Chiral
Algebra of 2D Gravity Coupled to $c\le 1$ Matter,'' preprint PRA-LPTHE 92-4
(Jan., 1992). }

\REF\KACHA{S. Kachru, ``Extra States and Symmetries in $D<2$ Closed String
Theory,'' preprint PUPT-1314 (April, 1992). }

\REF\VER{E. Verlinde, ``The Master Equation of 2D String Theory,''
preprint IASSNS-HEP-92/5 (Feb., 1992). }

As shown in several recent papers [\KACH, \PAS, \BER], the symmetry
algebra and transformation rules can be used to derive Ward identities
relating different correlation functions. In fact the sub-algebra of
charges with zero ghost number can be exploited to completely
determine the multi-tachyon amplitudes. Similarly one expects
that the amplitudes containing more than one discrete states may be
determined from amplitudes containing less number of discrete states,
by exploiting Ward identities derived from charges with non-zero
ghost number. Let us close this paper with some remarks about this
problem.

First, if one tries to compute the four-point amplitudes containing
the exotic discrete states ($O_{jm}$ or $P_{jm}$) directly,
one always gets zero or $\infty$. This is because the
cosmological constant has been taken to be $\mu=0$ and the momenta been taken
at discrete values. One can circumvent
the problems by taking a generic $\mu$ and compute the amplitudes in
the presence of a cosmological constant [\LI] and regularizing the
$D=2$ string theory by $D<2$
string theory or equivalently the $c<1$ matter coupled to gravity
[\FRA]. At least in certain cases, taking the limit $c\to 1 $ could
lead to some meaningful results.  In principle there seems no problem
to extend our approach to $D<2$ string theory, but in practice this
extension may be confronted with some serious technical problems. (See
refs. [\CHA, \DOTB, \KACHA].)

Second, while it is not too difficult to derive the Ward identities
[\VER, \PAS], it is highly non-trivial to use them in computing
correlation functions. The difficulty lies in choosing the proper charges
to derive a proper set of Ward identities that is soluble to give
more-point amplitude in terms of less-point amplitudes.  There seems
no guarantee that all the more-point amplitudes can be obtained from
less-point amplitudes by Ward identities [\VER].
Combined with the regularization
problem mentioned above, the practical use of Ward identities in computing
more correlation functions remains a topic to be attacked.

\vglue .6cm

\centerline{Acknowledgement}

The authors would like to thank Dr. Miao Li for interesting discussions.
The work was supported in part by U.S. NSF through grant No. PHY-9008452
and the work of Zhu at SISSA/ISAS was supported by an INFN post-doctoral
fellowship.

\vfill\eject

\centerline{\bf  Appendix A.  Schur Polynomials}

The elementary Schur polynomials $S_k(x)$ are defined through the generating
function
$$ \sum_{k\ge 0} S_k(x)z^k = \exp\{ \sum_{k\ge 1}  x_k  z^k\}. \eqn\apa$$
where the $x_k$'s, with $k=1$, 2, $\cdots$, are the arguments of the
polynomials. What we are using in this paper is the Schur polynomials having
$x_k$'s to be
$$ x_k = - i \delta^-{ \partial^k X^+\over k!}
         - i \delta^+{ \partial^k X^-\over k!} , \eqn\apb$$
where $\delta^+$ and $\delta^-$ are two real numbers. To
simplify our notations we denote
$$ S_k(x) = S_k(-i \delta^-{ \partial^j X^+\over j!}
- i \delta^+{ \partial^j X^-\over j!}) \equiv
S_k(-i\delta^-X^+-i\delta^+X^-) .
\eqn\apc$$
These special Schur polynomials are related with the exponential function

\noindent $\exp\{iX^{\pm}(z+w)\}$ as follows
$$ \eqalign{ \sum_{k\ge 0} S_k(-i&\delta^-X^+(w)-i\delta^+X^-(w)) z^k
\cr & = \exp\{
-i \delta^-(X^+(z+w)-X^+(w)) -i \delta^+(X^-(z+w)- X^-(w) ) \} ,} \eqn\apd$$
or
$$ S_k(-i\delta^-X^+-i\delta^+X^-)  = { 1\over k!}
\partial^k\Big( \exp\{ -i \delta^-X^+ -i \delta^+X^- \}\Big).  \eqn\ape$$
We have been able to prove the following identities for OPE's containing
Schur polynomials:
$$ \partial X^+(z) :S_k(-i\delta X^-(w) ) : \sim
i\delta \sum_{j=1 }^k { 1\over (z-w)^{j+1} } :S_{k-j} (-i\delta X^-(w)):,
\eqn\apf$$
$$ : S_k(-i\delta X^-(z) ) :\partial X^+(w) \sim
i\delta\sum_{j=1}^k { (-1)^{j+1} \over (z-w)^{j+1} }
:S_{k-j}(-i\delta X^-(z)):, \eqn\apg$$
$$ \eqalign{ : S_k(-i\delta X^-(z) ) : & : e^{isX^+(w)}:\cr
&  \sim \sum_{n=1}^k
{ \Gamma({\delta s +n}) \over n! \Gamma({\delta s}) }{(-1)^n \over (z-w)^n }
:S_{k-n} (-i\delta X^-(z) ) e^{isX^+(w)  } : ,}  \eqn\api$$
$$ \eqalign{  : e^{isX^+(z)} : & : S_k(-i\delta X^-(w) ) : \cr
& \sim \sum_{n=1}^k  { \Gamma({\delta s +n}) \over n! \Gamma({\delta s}) }
{1 \over (z-w)^n } :S_{k-n} (-i\delta X^-(w) ) e^{isX^+(z)  } : .} \eqn\apj$$
These identities are crucial in our explicit determination of the structure
constants of the chiral BSRT cohomology ring.  Let us prove eqs. $\apf$ and
$\api$. The other two equations can be proved similarly.

Writing the Schur polynomial $S_k(-i\delta X^-(w))$ in a contour
integration form
$$S_k(-i\delta X^-(w)) = \oint_0[\dd Z] {1 \over Z^{k+1}}
e^{-i\delta\sum_{j=1}^\infty
{Z^j\over j!} \partial^jX^-(w) } , \eqn\appa$$
we have
$$\eqalign{
\partial X^+(z)& :S_k(-i\delta X^-(w) ) : =\oint_0[\dd Z]
{1 \over Z^{k+1}} \partial X^+(z)
 :e^{-i\delta\sum_{j=1}^{\infty} { Z^j \over j! } \partial^j X^-(w) } : \cr
& \sim - i\delta \oint_0[\dd Z]
{1 \over Z^{k+1}} \sum_{j=1}^{\infty} {Z^j\over j!} \langle
\partial X^+(z) \partial^j X^-(w) \rangle :e^{-i\delta\sum_{j=1}^{\infty}
{ Z^j \over j! } \partial^j X^-(w) } :\cr
& = i\delta \oint_0[\dd Z]
{1 \over Z^{k+1}} \sum_{j=1}^{\infty} {Z^j\over (z-w)^{j+1} }
:e^{-i\delta\sum_{j=1}^{\infty}
{ Z^j \over j! } \partial^j X^-(w) } :\cr
& = i\delta \sum_{j=1}^{k}{1\over (z-w)^{j+1}} :S_{k-j}(-i\delta X^-(w) ). }
\eqn\appb$$
This proves $\apf $. To prove $\api$ we use the following equation for
the normal ordering of two exponential functions:
$$ :e^{-i\delta \partial^j X^-(z)}: :e^{isX^+(w)}:  \sim
e^{\delta s \langle \partial^j X^-(z) X^+(w) \rangle }
: e^{-i\delta \partial^j X^-(z) + isX^+(w) } : . \eqn\appc$$
We have
$$\eqalign{
 :S_k&(-i\delta X^-(w) ) ::e^{isX^+(w)}: =
\oint_0[\dd Z] { 1\over Z^{k+1} }
:e^{-i\delta\sum_{j=1}^{\infty} { Z^j \over j! } \partial^j X^-(w) } :
:e^{isX^+(w)}: \cr
& \sim  \oint_0[\dd Z] { 1\over Z^{k+1} }
e^{\delta s \sum_{j=1}^{\infty} { Z^j\over j!}
\langle \partial^j X^-(z) X^+(w) \rangle }
:e^{ -i\delta \partial^j X^-(z) + isX^+(w) } : \cr
& = \oint_0[\dd Z] { 1\over Z^{k+1} }
{ 1\over  ( 1+ { Z\over z-w} )^{\delta s} }
:e^{ -i\delta \partial^j X^-(z) + isX^+(w) } :  \cr
& =  \sum_{n=1}^k
{ \Gamma({\delta s +n}) \over n! \Gamma({\delta s}) }{(-1)^n \over (z-w)^n }
:S_{k-n} (-i\delta X^-(z) ) e^{isX^+(w)  } : . } \eqn\cccccc$$

The other formula we use in the text is [\BMP]
$$ \sum_{m=j+1 } ^k (m-j)x_{m-j}S_{k-m}(x) = (k-j) S_{k-j}(x) ,  \eqn\apk$$
which can easily be proved through the generating function:
$$ \eqalign{
\sum_{k\ge j} (k-j)S_{k-j}z^{k-j} =&  z{ \dd\over \dd x} \Big( \sum_{k\ge j}
S_{k-j}(x)z ^{k-j} \Big) \cr
 = & z{ \dd\over \dd z} \big( \exp (\sum x_kz^k) \big) = \sum_{k\ge 1} kx_k
z^k\exp(\sum_{k}x_kz^k ) \cr
= \sum_{k\ge 1} \sum_{l\ge 0} kx_k & S_l(z) z^{k+l} = \sum_{k\ge j } \Big(
\sum_{m= j+1}^k (m-j)x_{m-j} S_{k-m}(x) \Big) z^{k-j} . } $$

\vglue .5cm

\centerline{\bf  Appendix B. The Chiral Transformation Rules
and Chiral Charge Algebra}

\def\jja{j_1m_1}
\def\jjb{j_2m_2}
\def\cjja{j_1+j_2,m_1+m_2}
\def\cjjb{j_1+j_2-1,m_1+m_2}
\def\cjjc{j_2-j_1, m_1+m_2}
\def\cjjd{j_2-j_1+1, m_1+m_2}
\def\cjje{j_1-j_2, m_1+m_2}
\def\cjjf{j_1-j_2+1, m_1+m_2}
\def\jjaa{(m_1j_2-m_2j_1)}
\def\jjab{(m_1j_2+m_2j_1+m_1)}
\def\jjac{(m_1j_2+m_2j_1+m_2)}

In this appendix we present the full set of chiral transformation rules.
All of them are derived from  $\rahh$ and $\zza$ by using our results
on the cohomology ring structure. We have
$$ \eqalign{
\{ \hat{X}_{j_1m_1} , Y^+_{j_2m_2} \} =&
-(m_1j_2- m_2j_1)O_{j_1+j_2-1, m_1+m_2}, \cr
[\hat{X}_{j_1m_1}, P_{j_2m_2}]
=& -(m_1j_2+m_2j_1+m_1)Y^-_{j_2-j_1+1, m_1+m_2}, \cr
} $$
$$\eqalign{
\{\hat{X}_{j_1m_1} , aO_{j_2m_2} \} =& j_1O_{j_1+j_2,m_1+m_2}, \cr
[ \hat{X}_{j_1m_1} , aY^+_{j_2m_2}]  =&{j_1j_2\over j_1+j_2}Y^+_{j_1+j_2,
m_1+m_2}+{j_2\over j_1+j_2}(m_1j_2-m_2j_1)aO_{j_1+j_2-1, m_1+m_2}, \cr
[ \hat{X}_{j_1m_1} , aY^-_{j_2m_2}] =& j_1Y^-_{j_2-j_1, m_1+m_2}, \cr
\{\hat{X}_{j_1m_1}, aP_{j_2m_2}\} =& {j_1(j_2+1)\over j_2-j_1+1}
P_{j_2-j_1, m_1+m_2}\cr
& +{j_2+1\over j_2-j_1+1}(m_1j_2+m_2j_1+m_1)aY^-_{j_2-j_1+1, m_1+m_2}, \cr
} $$
$$\eqalign{
 [Q^+_{j_1m_1}, O_{j_2m_2}] =& (m_1j_2-m_2j_1)O_{j_1+j_2-1, m_1+m_2}, \cr
[Q^+_{j_1m_1}, Y^+_{j_2m_2} ] = &(m_1j_2-m_2j_1)Y^+_{j_1+j_2-1, m_1+m_2}, \cr
 [Q^+_{j_1m_1}, Y^-_{j_2m_2} ]
=& -(m_1j_2+m_2j_1+m_1)Y^-_{j_2-j_1+1, m_1+m_2} , \cr
 [Q^+_{j_1m_1}, P_{j_2m_2}]
=& -(m_1j_2+m_2j_1+m_1)P_{j_2-j_1+1, m_1+m_2}, \cr
} $$
$$\eqalign{
[Q^+_{\jja}, aO_{\jjb}] = & -{j_1(j_1-1)\over j_1+j_2} Y^+_{\cjja}
 \cr & + {j_2+1\over j_1+j_2} \jjaa aO_{\cjjb}, \cr
   [Q^+_{\jja}, aY^+_{\jjb}] =& {j_2\over j_1+j_2-1} \jjaa aY^+_{\cjjb}, \cr
   [Q^+_{\jja}, aY^-_{\jjb}] =& -{j_1(j_1-1)\over j_2-j_1+1} P_{\cjjc}
 \cr & -{j_2\over j_2-j_1+1}\jjab aY^-_{\cjjd}, \cr
   [Q^+_{\jja}, aP_{\jjb}] =& -{j_2+1\over j_2-j_1+2}\jjab aP_{\cjjd}, \cr
} $$
$$\eqalign{
   [Q^-_{\jja}, Y^+_{\jjb}] =& \jjac Y^-_{\cjjf}, \cr
   [Q^-_{\jja}, aO_{\jjb}] =& (j_1+1)Y^-_{\cjje}, \cr
   [Q^-_{\jja}, aY^+_{\jjb}] =& -{j_2(j_1+1)\over j_1-j_2+1} P_{\cjje}
 \cr & -{j_2\over j_1-j_2+1}\jjac aY^-_{\cjjf}, \cr
} $$
$$\eqalign{
   [\hat{Z}_{\jja}, O_{\jjb}] =& -\jjac Y^-_{\cjjf}, \cr
   \{\hat{Z}_{\jja}, Y^+_{\jjb}\} =& -\jjac P_{\cjjf}, \cr
} $$
$$\eqalign{
   \{\hat{Z}_{\jja}, aO_{\jjb}\} =& -{(j_1+1)(j_1+2)\over j_1-j_2+1}P_{\cjje}
 \cr & -{j_2+1\over j_1-j_2+1}\jjac aY^-_{\cjjf}, \cr
   [\hat{Z}_{\jja}, aY^+_{\jjb}] =& -{j_2\over j_1-j_2+2}\jjac aP_{\cjjf}, \cr
} $$
$$\eqalign{
   [a\hat{X}_{\jja}, O_{\jjb}] =& j_2O_{\cjja}, \cr
   [a\hat{X}_{\jja}, Y^+_{\jjb}] =& {j_2(j_2-1)\over j_1+j_2} Y^+_{\cjja}
 \cr & +{j_1+1\over j_1+j_2}\jjaa aO_{\cjjb}, \cr
   [a\hat{X}_{\jja}, Y^-_{\jjb}] =& -(j_2+1)Y^-_{\cjjc}, \cr
   [a\hat{X}_{\jja}, P_{\jjb}] =& -{(j_2+1)(j_2+2)\over j_2-j_1+1} P_{\cjjc}
 \cr & -{j_1+1\over j_2-j_1+1}\jjab aY^-_{\cjjd}, \cr
} $$
$$\eqalign{
   [a\hat{X}_{\jja}, aO_{\jjb}] =& -(j_1-j_2)aO_{\cjja}, \cr
   [a\hat{X}_{\jja}, aY^+_{\jjb}] =& -{j_2(j_1-j_2+1)\over j_1+j_2}
aY^+_{\cjja}, \cr
   [a\hat{X}_{\jja}, aY^-_{\jjb}] =& -(j_1+j_2+1)aY^-_{\cjja}, \cr
   [a\hat{X}_{\jja}, aP_{\jjb}] =& -{(j_2+1)(j_1+j_2+2)\over j_2-j_1+1}
aP_{\cjjc}, \cr
} $$
$$\eqalign{
   [aQ^+_{\jja}, O_{\jjb}] =& {j_1j_2\over j_1+j_2}Y^+_{\cjja}
 -{j_1\over j_1+j_2} \jjaa aO_{\cjjb}, \cr
   \{aQ^+_{\jja}, Y^+_{\jjb}\} =& -{j_1\over j_1+j_2-1}\jjaa aY^+_{\cjjb}, \cr
   \{aQ^+_{\jja}, Y^-_{\jjb}\} =& { j_1(j_2+1) \over j_2-j_1+1}P_{\cjjc}
 \cr & -{j_1\over j_2-j_1+1}\jjab aY^-_{\cjjd}, \cr
   [aQ^+_{\jja}, P_{\jjb}] =& -{j_1\over j_2-j_1+2} \jjab aP_{\cjjd}, \cr
} $$
$$\eqalign{
   \{aQ^+_{\jja}, aO_{\jjb}\} =& {j_1(j_1-j_2-1)\over j_1+j_2}
aY^+_{\cjja}, \cr
   [aQ^+_{\jja}, aY^-_{\jjb}] =& {j_1(j_1+j_2)\over j_2-j_1+1} aP_{\cjjc}, \cr
} $$
$$\eqalign{
   [aQ^-_{\jja}, O_{\jjb}] =& j_2Y^-_{\cjje}, \cr
   \{aQ^-_{\jja}, Y^+_{\jjb}\} =& -{j_2(j_2-1)\over j_1-j_2+1} P_{\cjje}
 \cr & -{j_1\over j_1-j_2+1} \jjac aY^-_{\cjjf}, \cr
   \{aQ^-_{\jja}, aO_{\jjb}\} =& -(j_1+j_2+1)aY^-_{\cjje}, \cr
   [aQ^-_{\jja}, aY^+_{\jjb}] =& {j_2(j_1+j_2)\over j_1-j_2+1} aP_{\cjje}, \cr
} $$
$$\eqalign{
   [a\hat{Z}_{\jja}, O_{\jjb}] =& {j_2(j_1+1)\over j_1-j_2+1} P_{\cjje}
 \cr & +{j_1+1\over j_1-j_2+1} \jjac aY^-_{\cjjf},  \cr
   [a\hat{Z}_{\jja}, Y^+_{\jjb}] =& {j_1+1\over j_1-j_2+1} \jjac
aP_{\cjjf}, \cr
   [a\hat{Z}_{\jja}, aO_{\jjb}] =& {(j_1+1)(j_1+j_2+2)\over j_1-j_2+1}
aP_{\cjje}, \cr
}$$
$$\eqalign{
[\hat{X}_{j_1m_1} , O_{j_2m_2} ] =& 0, \cr
   [Q^-_{\jja}, O_{\jjb}] =& 0, \cr
   [Q^-_{\jja}, P_{\jjb}] =& 0, \cr
   [Q^-_{\jja}, aP_{\jjb}] =&0, \cr
   [aQ^-_{\jja}, P_{\jjb}] =& 0, \cr
   [\hat{Z}_{\jja}, P_{\jjb}] =& 0, \cr
   \{\hat{Z}_{\jja}, aP_{\jjb}\} =& 0, \cr
   \{aQ^+_{\jja}, aP_{\jjb}\} =& 0, \cr
   \{aQ^-_{\jja}, aP_{\jjb}\} =& 0, \cr
   [a\hat{Z}_{\jja}, P_{\jjb}] =& 0, \cr
   [a\hat{Z}_{\jja}, aY^-_{\jjb}] =& 0, \cr }\qquad
\eqalign{
\{ \hat{X}_{j_1m_1} , Y^-_{j_2m_2}\} =& 0, \cr
   [Q^-_{\jja}, Y^-_{\jjb}] =& 0, \cr
   [Q^-_{\jja}, aY^-_{\jjb}] =& 0, \cr
   \{aQ^-_{\jja}, Y^-_{\jjb}\} =& 0, \cr
   \{\hat{Z}_{\jja}, Y^-_{\jjb}\} =& 0, \cr
   [\hat{Z}_{\jja}, aY^-_{\jjb}] =& 0, \cr
   [aQ^+_{\jja}, aY^+_{\jjb}] =& 0, \cr
   [aQ^-_{\jja}, aY^-_{\jjb}] =& 0, \cr
   [a\hat{Z}_{\jja}, Y^-_{\jjb}] =& 0, \cr
   [a\hat{Z}_{\jja}, aY^+_{\jjb}] =& 0, \cr
   [a\hat{Z}_{\jja}, aP_{\jjb}] =& 0. \cr } $$
The chiral charge algebra can be obtained from the above equations by acting
with $\oint_0[\dd w]$  $\oint_w[\dd z] b(z)$ from the left. This action
transforms all the local fields into the corresponding charges, as shown
in sec. 5.

\vglue .5cm

\centerline{ \bf Appendix C. Proof of Eqs. $\pqq$ and $\pyqq$ }

In this appendix we will give some details about the computation of the
transformation rules in closed string theory and also the proof of eqs.
$\pqq$ and $\pyqq$. To simplify the presentation we omit the ``magnetic''
index of the (chiral) discrete states and write the chiral cohomology ring
structure equation as follows:
$$V_{j_1}V_{j_2} = A_j + aB_{j'} , \eqn\cappa$$
where $V_{j_1}$, $ V_{j_2}$, $A_j$ and $B_{j'}$ are all chiral relative
discrete states. By using $\rahh$ and $\zza$ we can write the chiral
transformation rules as follows:
$$ \eqalign{
& [\hat{V}_{j_1}, V_{j_2}] = g_B(j')B_{j'}, \cr
& [\hat{V}_{j_1}, aV_{j_2}]
= (-1)^{[V_{j_1}]} \big( (g_V(j_1) -1)A_j -g_V(j_2)aB_{j'} \big), \cr
& [a\hat{V}_{j_1}, V_{j_2}] = (g_V(j_2) - 1)A_j - g_V(j_1)aB_{j'}, \cr
& [a\hat{V}_{j_1}, aV_{j_2}]
= -(-1)^{[V_{j_1}]} \big( g_V(j_1) - g_V(j_2) \big) aA_j.
} \eqn\cappb$$
Note that the relation $g_A(j)= g_B(j')= g_V(j_1)+g_V(j_2)-1$ has been used
in our derivation of the above rules.

A relative closed string discrete state is just the product a left relative
discrete state and a right relative discrete states: $V_{j_1\bar{j}_1} =
V_{j_1}\bar{V}_{\bar{j}_1 } $.  A necessary and sufficient condition for the
pairing (at the $SU(2)$ point, of course) is $g_V(j_1)=g_{\bar{V}}(\bar{j}_1)$.
The charge $\hat{V}_{j_1\bar{j}_1}$ derived from the discrete state
$V_{j_1\bar{j}_1}$ can be written symbolically  (for the purpose of doing
calculation) as follows
$$\hat{V}_{j_1\bar{j}_1} = \hat{V}_{j_1} \bar{V}_{\bar{j}_1} -
(-1)^{[V_{j_1}]} {V}_{j_1} \hat{\bar{V}}_{\bar{j}_1} . \eqn\cappc$$
The action of $\hat{V}_{j_1\bar{j}_1}$ on $V_{j_2\bar{j}_2} =
V_{j_2}\bar{V}_{\bar{j}_2 } $ is computed as follows:
$$\eqalign{
[ \hat{V}_{j_1\bar{j}_1} &,V_{j_2\bar{j}_2} ] =
  [ \hat{V}_{j_1} \bar{V}_{\bar{j}_1} -
(-1)^{[V_{j_1}]} {V}_{j_1} \hat{\bar{V}}_{\bar{j}_1},
V_{j_2}\bar{V}_{\bar{j}_2 }] \cr
= & (-1)^{[\bar{V}_{\bar{j}_1} ][V_{j_2}]} \Big( [ \hat{V}_{j_1}, V_{j_2}]
\bar{V}_{\bar{j}_1} \bar{V}_{\bar{j}_2} - (-1)^{[V_{j_1}] + [V_{j_2}] }
V_{j_1}V_{j_2}[\hat{\bar{V}}_{\bar{j}_1}, \bar{V}_{\bar{j}_2} ]\Big)  \cr
= & (-1)^{[\bar{V}_{\bar{j}_1} ][V_{j_2}] } g_B(j')
\Big(B_{j'}\bar{A}_{\bar{j}}
- (-1)^{[V_{j_1}] + [V_{j_2}] } A_j \bar{B}_{\bar{j}'} \cr
& \qquad\qquad\qquad \quad
- (-1)^{[V_{j_1}] + [V_{j_2}] } (a+\bar{a}) B_{j`}\bar{B}_{\bar{j}'} \Big) .
} \eqn\cappd$$
 From the above result we see that the action of any charge on the closed
string discrete state gives only physical closed string states. No states
other than  those listed in $\zze$ and $\zzf$ are created.

The action of $\hat{V}_{j_1\bar{j}_1}$ on the absolute state $(a+\bar{a})
V_{j_2\bar{j}_2}$ can also be computed. We have
$$ \eqalign{
[ \hat{V}_{j_1\bar{j}_1} &,(a+\bar{a})V_{j_2\bar{j}_2} ] =
[ \hat{V}_{j_1} \bar{V}_{\bar{j}_1} -
(-1)^{[V_{j_1}]} {V}_{j_1} \hat{\bar{V}}_{\bar{j}_1},
aV_{j_2}\bar{V}_{\bar{j}_2 }+(-1)^{[V_{j_2}]}
V_{j_2}\bar{a}\bar{V}_{\bar{j}_2 }] \cr
= & (-1)^{[\bar{V}_{\bar{j}_1} ][V_{j_2}] + [\bar{V}_{\bar{j}_1} ]
+ [V_{j_1}] -1} g_B(j')(a+\bar{a}) \Big(B_{j'}\bar{A}_{\bar{j}}
-   (-1)^{[V_{j_1}] + [V_{j_2}] } A_j \bar{B}_{\bar{j}'}  \Big) \cr
= & (-1)^{ [\bar{V}_{\bar{j}_1} ]
+ [V_{j_1}] -1} (a+\bar{a})
[ \hat{V}_{j_1\bar{j}_1} ,V_{j_2\bar{j}_2} ],  \cr
} \eqn\cappe$$
by using $\cappd$. This proves $\pqq$.

Eq. $\pyqq$ can be proved similarly. We have
$$ \eqalign{
[ (a+\bar{a})&\hat{V}_{j_1\bar{j}_1} ,V_{j_2\bar{j}_2} ] \cr
= & [ a\hat{V}_{j_1} \bar{V}_{\bar{j}_1}
 - {V}_{j_1}\bar{a} \hat{\bar{V}}_{\bar{j}_1}
 + (-1)^{[V_{j_1}]} a{V}_{j_1} \hat{\bar{V}}_{\bar{j}_1}
 + (-1)^{[V_{j_1}]} \hat{V}_{j_1}\bar{a} \bar{V}_{\bar{j}_1},
V_{j_2}\bar{V}_{\bar{j}_2 }] \cr
= & (-1)^{[\bar{V}_{\bar{j}_1} ][V_{j_2}] }
g_A(j)(a+\bar{a}) \Big(-B_{j'}\bar{A}_{\bar{j}}
 +   (-1)^{[V_{j_1}] + [V_{j_2}] } A_j \bar{B}_{\bar{j}'}  \Big) \cr
= & - (a+\bar{a})
[ \hat{V}_{j_1\bar{j}_1} ,V_{j_2\bar{j}_2} ],  \cr
} \eqn\cappf$$
and
$$ \eqalign{
[ (a+\bar{a})\hat{V}_{j_1\bar{j}_1},(a+\bar{a})V_{j_2\bar{j}_2} ] = &
[ a\hat{V}_{j_1} \bar{V}_{\bar{j}_1}
 - {V}_{j_1}\bar{a} \hat{\bar{V}}_{\bar{j}_1}
 + (-1)^{[V_{j_1}]} a{V}_{j_1} \hat{\bar{V}}_{\bar{j}_1}\cr
& +  (-1)^{[V_{j_1}]} \hat{V}_{j_1}\bar{a} \bar{V}_{\bar{j}_1},
aV_{j_2}\bar{V}_{\bar{j}_2 }+(-1)^{[V_{j_2}]}
V_{j_2}\bar{a}\bar{V}_{\bar{j}_2 }] \cr
= & \quad 0.  \cr
}\eqn\cappg$$
This concludes the proof of $\pyqq$.

\vfill\eject

\refout

\bye